\documentclass[aps,10pt,nofootinbib,groupedaddress,preprintnumbers,superscriptaddress,twocolumn
]{revtex4}
\usepackage[dvipdfmx]{graphicx} 

\usepackage{amssymb,amsfonts,amsmath,bm,color,multirow,cases,empheq,hyperref}
\usepackage{here}
\usepackage{mathrsfs}

\usepackage{enumerate}
\usepackage{ulem}
\usepackage{afterpage}

\hypersetup{%
 setpagesize=false,
 bookmarksnumbered=true,%
 bookmarksopen=true,%
 colorlinks=true,%
 linkcolor=blue,%
 citecolor=red}

\def\bs{b^{\mathrm{s}}}

\def\bSPO{b_{\mathrm{SPO}}}
\def\bm{b_{\mathrm{m}}}
\def\d{\mathrm{d}}

\def\pa{\partial}
\def\qSPO{q_{\mathrm{SPO}}}

\def\qmax{q_{\mathrm{max}}}
\def\qmin{q_{\mathrm{min}}}
\def\qs{q^{\mathrm{s}}}
\def\qt{q^{\mathrm{t}}}
\def\qb{\bar{q}}
\def\rh{r_{\mathrm{H}}}
\def\rc{r^{\mathrm{c}}}
\def\s{\sigma}

\def\tm{\theta_{\mathrm{m}}}

\def\vp{\varphi}
\def\<{\langle}
\def\>{\rangle}
\def\={\equiv}
\def\2{I\hspace{-.1em}I}
\def\3{I\hspace{-.1em}I\hspace{-.1em}I}
\def\4{I\hspace{-.1em}V}

\begin{document}

\title{Photon escape in the extremal Kerr black hole spacetime}

\author{Kota Ogasawara}
\email{kota@tap.scphys.kyoto-u.ac.jp}
\affiliation{Theoretical Astrophysics Group, Department of Physics, Kyoto University, Kyoto 606-8502, Japan}

\author{Takahisa Igata}
\email{igata@post.kek.jp}
\affiliation{KEK Theory Center, Institute of Particle and Nuclear Studies, High Energy Accelerator Research Organization, Tsukuba 305-0801, Japan}

\date{\today}
\preprint{KUNS-2900}
\preprint{KEK-Cosmo-0279}
\preprint{KEK-TH-2363}

\begin{abstract}
We consider necessary and sufficient conditions for photons emitted from an arbitrary spacetime position of the extremal Kerr black hole to escape to infinity.
The radial equation of motion determines the necessary conditions for photons emitted from $r=r_*$ to escape to infinity, and the polar angle equation of motion further restricts the allowed region of photon motion.
From these two conditions, we provide a method to visualize a two-dimensional photon impact parameter space that allows photons to escape to infinity, i.e., the escapable region.
Finally, we completely identify the escapable region for the extremal Kerr black hole spacetime.
This study has generalized our previous result 
[K.~Ogasawara and T.~Igata, Phys. Rev. D \textbf{103}, 044029 (2021)], which focused only on light sources near the horizon, to the classification covering light sources in the entire region.
\end{abstract}

\maketitle

\section{Introduction}

In recent years, the observation of the vicinity of a black hole has made great progress. A bright ring structure and associated shadow of the M87 galactic center were discovered in 2019 by the Event Horizon Telescope Collaboration~\cite{Akiyama:2019cqa}. This result suggests that the central object is a supermassive black hole. However, the possibility that the central object is a horizonless compact object has not yet been dismissed~\cite{Cardoso:2019rvt}.
In general, the difference between a black hole and a black hole candidate will be noticeable in phenomena near the horizon. Therefore, it is important to detect signals, i.e., photons, coming from the vicinity of the horizon radius of a central object and identify them uniquely and accurately.
As the observation progresses in the future, we will be able to clarify various properties of central objects.
The black hole observations, including the shadow observations, require capturing photons that have passed near the horizon radius, shaken off the strong gravitational field, and finally escaped to infinity.
Therefore, how often photons can escape from the light source to infinity, that is, the escape probability, is an important issue.

The escape of photons was first revealed by Synge, who estimated photon escape cones in the Schwarzschild black hole~\cite{Synge:1966okc}. He found that 50\% of photons emitted from the photon sphere could escape to infinity, while the remaining 50\% were trapped by the black hole. Furthermore, the opening angle of the escape cone becomes smaller as the photon emission point approaches the horizon, and eventually, it becomes zero in the horizon limit. This implies that the observability of the vicinity of the horizon is extremely low, and it seems quite natural considering the nature of the black hole, from which nothing can escape.

However, it has recently been reported that photons emitted from the vicinity of a rapidly rotating black hole can have a large escape probability.
In our previous work, we showed that 29.1\% of photons could escape to infinity, even when a uniform emitter at rest in a locally nonrotating frame arbitrarily approaches the extremal Kerr horizon~\cite{Ogasawara:2019mir}. 
For the subextremal case, the escape probability becomes zero in the same limit, but for the near-extremal case, it is maintained at about 30\% until just before the horizon.
These results imply that the vicinity of a rapidly rotating black hole is more visible than that of a slowly rotating one.
The escapes of photons in other black hole spacetimes were discussed in Refs.~\cite{Semerak:1996,Stuchlik:2018qyz,Zhang:2020pay}, and the ratio of photons trapped by a black hole was discussed in Ref.~\cite{Takahashi:2010ai}.

More recently, the escape probability of photons emitted from an emitter in a stable circular orbit of a Kerr black hole was shown to be more than 50\% for an arbitrary spin parameter and an arbitrary orbital radius~\cite{Igata:2019hkz,Gates:2020sdh,Gates:2020els}.
Furthermore, the Doppler blueshift overcomes the gravitational redshift according to the direction of photon emission with respect to the direction of source motion, so that photons can reach a distant observer with an observable frequency band.
These two effects are due to the relativistic boost or beaming caused by the proper motion of the emitter, and in recent years, such relativistic effects have been actively discussed~\cite{GRAVITY:2018ofz,Saida:2019mcz,Iwata:2020pka,Igata:2021njn}.
The analytic value of the escape probability and Doppler blueshift of various emitters were recently found by using the near-horizon geometry of the extremal Kerr black hole~\cite{Gates:2020els,Yan:2021yuo,Yan:2021ygy}.

The previous works of photon escape have considered the source confined to the equatorial plane.
However, if a small perturbation is applied to the source orbiting around a Kerr black hole, it will no longer be confined to the equatorial plane and will fall into the black hole.
A thorough analysis of such a nonequatorial plane emission of photons will be necessary for black hole observations which are expected to develop further in the future.

The purpose of this paper is to completely classify the necessary and sufficient parameter region for photons emitted from an arbitrary spacetime position of the extremal Kerr black hole to escape to infinity.
This study generalizes the previous result~\cite{Ogasawara:2020frt}, which focused only on light sources near the horizon, to a classification that covers light sources in the entire region. 

This paper is organized as follows.
In Sec.~\ref{sec:2}, we consider the equations of a photon, i.e., the null geodesic equations, in the Kerr black hole spacetime. In Sec.~\ref{sec:3}, we clarify the necessary and sufficient conditions for photons to escape from an arbitrary spacetime position to infinity by using the allowed region of motion and the spherical photon orbits (SPOs). In addition, we develop a method to visualize a two-dimensional photon impact parameter space that allows photons to escape to infinity.
In Sec.~\ref{sec:4}, we introduce critical polar angles and critical values of an impact parameter to specify the escapable region explicitly. Using the visualization method and critical values, we completely evaluate the escapable region in Sec.~\ref{sec:5}.
Section \ref{sec:6} is devoted to discussion.
In this paper, we use units in which $c=1$ and $G=1$.

\section{General Null Geodesic in the Kerr Black Hole Spacetime}
\label{sec:2}
We review the general null geodesic in the Kerr black hole spacetime. The Kerr metric in the Boyer-Lindquist coordinates is given by
\begin{align}
g_{\mu\nu}\d x^\mu\d x^\nu=&-\frac{\Sigma \Delta}{A}\d t^2+\frac{\Sigma}{\Delta}\d r^2+\Sigma\d \theta^2 \nonumber\\
&+\frac{A}{\Sigma}\sin^2\theta\left(\d\vp-\frac{2Mar}{A}\d t\right)^2,
\end{align}
\begin{align}
\Sigma&=r^2+a^2\cos^2\theta,~~\Delta=r^2-2Mr+a^2, \nonumber\\
A&=\left(r^2+a^2\right)^2-a^2\Delta^2\sin^2\theta,
\end{align}
where $M$ and $a$ denote the mass and spin parameters, respectively. The spacetime is stationary and axisymmetric with two corresponding Killing vectors $\xi^a\pa_a=\pa_t$ and $\psi^a\pa_a=\pa_\vp$, respectively. Furthermore, the spacetime has the Killing tensor $K_{ab}$ defined by \cite{Walker:1970un}
\begin{align}
K_{ab}=&\;\Sigma^2(\d\theta)_{a}(\d\theta)_{b}+\sin^2\theta\left[\left(r^2+a^2\right)(\d\vp)_{a}-a(\d t)_{a}\right] \nonumber\\
&\times \left[\left(r^2+a^2\right)(\d\vp)_{b}-a(\d t)_{b}\right]-a^2\cos^2\theta g_{ab}.
\end{align}
We adopt units in which $M=1$ in what follows.

Let us consider null geodesics with 4-momentum $k^a$. According to the existence of $\xi^a$, $\psi^a$, and $K_{ab}$, a photon has three constants of motion \cite{Carter:1963}
\begin{align}
E&=-\xi^ak_a=-k_t,~~L=\psi^ak_a=k_\vp, \nonumber\\
Q&=K_{ab}k^ak^b-(L-aE)^2,
\end{align}
where $E$, $L$, and $Q$ are the conserved energy, angular momentum, and Carter constant, respectively. Since we consider only a photon escaping to infinity, we assume that $E>0$. Introducing impact parameters
\begin{align}
b=\frac{L}{E},~~q=\frac{Q}{E^2},
\end{align}
and rescaling $k^a$ as $k^a/E\to k^a$, we obtain the null geodesic equations parametrized by $(b,q)$:
\begin{align}
\Sigma \dot{t}&=a\left(b-a\sin^2\theta\right)+\frac{r^2+a^2}{\Delta}\left(r^2+a^2-ab\right),\\
\Sigma \dot{\vp}&=\frac{b-a\sin^2\theta}{\sin^2\theta}+\frac{a}{\Delta}\left(r^2+a^2-ab\right),\\
\Sigma \dot{r}&=\s_r\sqrt{R},\\
\Sigma \dot{\theta}&=\s_\theta\sqrt{\Theta},
\end{align}
where the dots denote derivatives with respect to an affine parameter, $\s_r=\mathrm{sgn}(\dot{r})$, $\s_\theta=\mathrm{sgn}(\dot{\theta})$, and
\begin{align}
R&=\left(r^2+a^2-ab\right)^2-\Delta\left[q+(b-a)^2\right],\\
\Theta&=q-\cot^2\theta\left(b^2-a^2\sin^2\theta\right). \label{def:Theta}
\end{align}
The allowed region for photon motion is $R\geq0$ and $\Theta\geq0$. From now on, we focus on the extremal Kerr black hole spacetime, i.e., $a=1$. Thus, the event horizon is located at $r=\rh=1$.

Let us clarify the allowed parameter region restricted by $R\geq0$. The function is factored as
\begin{align}
R=r(2-r)(b-b_1)(b-b_2),
\end{align}
where
\begin{align}
b_1(r;q)&=\frac{-2r+(r-1)\sqrt{r^4-r(r-2)q}}{r(r-2)},\\
b_2(r;q)&=\frac{-2r-(r-1)\sqrt{r^4-r(r-2)q}}{r(r-2)},
\end{align}
which denote the values of $b$ at the radial turning point. The allowed range of $b$ derived from $R\geq0$ is given by
\begin{align}
\begin{array}{lll}
b\leq b_1,~b\geq b_2 & \mathrm{for} & 1<r<2,\\
b_1<b<b_2 & \mathrm{for} & r>2.
\end{array}
\label{allowed_bR}
\end{align}
Note that $b_2$ is singular at $r=2$, but $R$ is finite there. We will not consider $b\geq b_2$ for $1<r<2$ because this range is for a negative energy photon, and such a photon cannot escape to infinity.

We also clarify the allowed parameter region restricted by $\Theta\geq0$. It reads
\begin{align}
q\geq\cot^2\theta\left(b^2-\sin^2\theta\right),
\end{align}
so that the allowed range of $b$ derived from $\Theta\geq0$ is given by
\begin{align}
-B\leq b\leq B,
\label{allowed_bTheta}
\end{align}
where
\begin{align}
B(\theta;q)=\tan\theta\sqrt{q+\cos^2\theta}.
\end{align}
Thus, the allowed region for photon motion is given by the common region of Eqs.~\eqref{allowed_bR} and \eqref{allowed_bTheta}.

\begin{figure}[t]
\centering
\includegraphics[width=7cm]{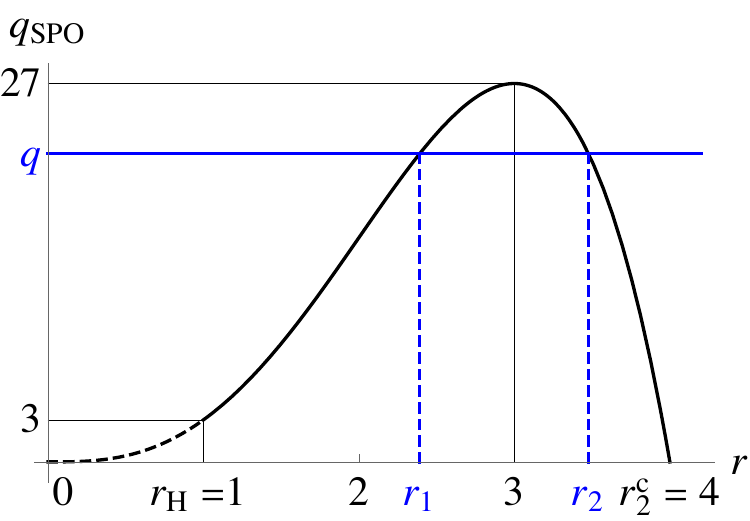}
\caption{Relation between $q$ and the radii $r_i$. The function $\qSPO(r)$ is shown by a black curve, which is solid outside the horizon and dashed inside it. The intersections of the blue solid line $q$ and the black solid curve $\qSPO(r)$ give the radii of SPOs, $r_1$ and $r_2$.}
\label{fig_qSPO}
\end{figure}

Next, we consider the extremum points of $b_i$ ($i=1,2$), which characterize the photon escape conditions. The photon orbits staying at the extrema, i.e., the orbits with $\dot{r}=0$ and $\ddot{r}=0$, are known as the SPOs~\cite{Teo:2003}. Solving the equivalent conditions, $R=0$ and $\d R/\d r=0$, we obtain $b$ and $q$ as functions of the SPO radius:
\begin{align}
b&=\bSPO(r)=-r^2+2r+1,\label{bSPO}\\
q&=\qSPO(r)=r^3(4-r).\label{qSPO}
\end{align}
Outside the horizon, $\qSPO(r)$ has a unique local maximum with the value $27$ at $r=3$. Eliminating $r$ from Eqs.~\eqref{bSPO} and \eqref{qSPO}, we obtain the extremum values as
\begin{align}
b=\bs_i(q)=\bSPO\big(r_i(q)\big),
\end{align}
where $r=r_i(q)$ ($r_1\leq r_2$) are the radii of SPOs and are the real solutions of $q=\qSPO(r)$. Note that $r_1$ ($r_2$) increases (decreases) monotonically with $q$ in the range
\begin{align}
r_1(0)=0\leq r_1(q)\leq3\leq r_2(q)\leq 4=\rc_2,
\end{align}
where $\rc_2=r_2(0)=4$ is the radius of the unstable photon circular orbit. The number of real roots of $q=\qSPO(r)$ outside the horizon depends on $q$. There exists a single root $r_2$ for $0\leq q\leq3$, while there exist two roots $r_1$ and $r_2$ for $3<q<27$. For $q=27$, $r_i$ coincide with each other at $r_1=r_2=3$, so that $\bs_i$ coincide with $\bs_1=\bs_2=-2$.
Figure \ref{fig_qSPO} shows a relation between $q$ and the radii $r_i$.

\section{Photon Escape Condition}
\label{sec:3}
We consider the escape condition of a photon emitted from an arbitrary spacetime position $(r,\theta)=(r_*,\theta_*)$. Since the Kerr black hole spacetime is reflection symmetric with respect to the equatorial plane $\theta=\pi/2$, we consider only the range $0<\theta_*<\pi/2$ in what follows. The cases of $\theta_*=0$ and $\theta_*=\pi/2$ will be considered in Appendix \ref{App:theta-0-pi}.

The necessary and sufficient conditions for photons to escape are that they have appropriate parameters to reach infinity from $r=r_*$ (necessary condition) and are in the allowed region determined by the variable $\theta_*$. In the following subsections, we consider the photon escape conditions for $q\geq0$ and $q<0$ separately.

\subsection{Necessary condition for photon escape, $q\geq0$}
\begin{figure*}[t]
\centering
\includegraphics[width=18cm]{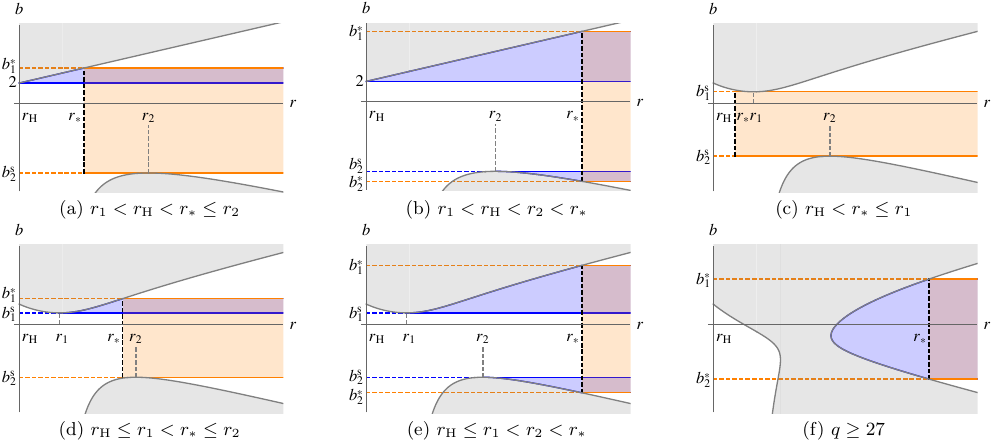}
\caption{Range of $b$ in which a photon satisfies a necessary condition for escape. Gray solid curves denote $b_i(r;q)$ as a function of $r$ with a fixed $q$, and black dashed vertical lines denote the emission point $r=r_*$. Gray regions denote forbidden regions, while the other regions denote allowed regions. Orange and blue regions represent the parameter ranges of $b$ where photons satisfy a necessary condition for escape with $\s_r=+$ and $\s_r=-$, respectively. (a) and (b) ($0\leq q<3$): the function $b_2$ has an extremum $\bs_2$ at $r=r_2$, but $b_1$ has no extremum. (c)--(e) ($3\leq q<27$): the functions $b_i$  have extrema $\bs_i$ at $r=r_i$, respectively. (f) ($q\geq27$): the allowed region is disconnected and the functions $b_i$ no longer have extrema.}
\label{fig_bi6cases}
\end{figure*}

Let us consider the behavior of $b_i(r;q)$ to determine the range of $b$ in which photons with $q\geq0$ satisfy the necessary condition to escape from $r=r_*$ to infinity. We can see the typical shape of $b_i$ as gray curves in Figs.~\ref{fig_bi6cases}(a) and \ref{fig_bi6cases}(b) for $0\leq q<3$, in Figs.~\ref{fig_bi6cases}(c)--\ref{fig_bi6cases}(e) for $3\leq q<27$, and in Fig.~\ref{fig_bi6cases}(f) for $q\geq27$. Gray regions denote forbidden regions of photon motion. Orange and blue regions represent the parameter range of $b$ where photons satisfy a necessary condition for escape with $\s_r=+$ and $\s_r=-$, respectively. 

In the case $r_1<\rh<r_*\leq r_2$, photons initially emitted outward (i.e., $\s_r=+$) with $\bs_2<b\leq b_1^*$ can escape [see orange region in Fig.~\ref{fig_bi6cases}(a)], and photons initially emitted inward (i.e., $\s_r=-$) with $2<b<b_1^*$ also can escape [see blue region in Fig.~\ref{fig_bi6cases}(a)], where
\begin{align}
b_i^*\=b_i(r_*;q),~~b_1(\rh;q)=2.
\end{align}

In the case $r_1<\rh<r_2<r_*$, photons initially emitted outward (i.e., $\s_r=+$) with $b_2^*\leq b\leq b_1^*$ can escape [see orange region in Fig.~\ref{fig_bi6cases}(b)], and photons initially emitted inward (i.e., $\s_r=-$) with $b_2^*<b<\bs_2$ or $2<b<b_1^*$ also can escape [see blue region in Fig.~\ref{fig_bi6cases}(b)].

In the case $\rh<r_*\leq r_1$, only photons initially emitted outward (i.e., $\s_r=+$) with $\bs_2<b<\bs_1$ can escape [see orange region in Fig.~\ref{fig_bi6cases}(c)].

In the case $\rh\leq r_1<r_*\leq r_2$, photons initially emitted outward (i.e., $\s_r=+$) with $\bs_2<b\leq b_1^*$ can escape [see orange region in Fig.~\ref{fig_bi6cases}(d)], and photons initially emitted inward (i.e., $\s_r=-$) with $\bs_1<b<b_1^*$ also can escape [see blue region in Fig.~\ref{fig_bi6cases}(d)].

In the case $\rh\leq r_1<r_2<r_*$, photons initially emitted outward (i.e., $\s_r=+$) with $b_2^*\leq b\leq b_1^*$ can escape [see orange region in Fig.~\ref{fig_bi6cases}(e)], and photons initially emitted inward (i.e., $\s_r=-$) with $b_2^*<b<\bs_2$ or $\bs_1<b<b_1^*$ also can escape [see blue region in Fig.~\ref{fig_bi6cases}(e)].

For $q\geq27$, the allowed region is disconnected. Therefore, if $r_*$ is in the outer allowed region, photons must have $b_2^*\leq b\leq b_1^*$ and always can escape [see orange and blue region in Fig.~\ref{fig_bi6cases}(f)].

\begin{table*}[t]
\centering
\caption{Necessary conditions for photon escape from $r=r_*$ to infinity. Here ``n/a" means not applicable.}
\begin{tabular}{lllll}
\hline\hline
Radial position of an emitter ~~& $q$ & $b$ ($\s_r=+$) & $b$ ($\s_r=-$) & Shape of $b_i$
\\ \hline
(i) $\rh<r_*<3$ & $0\leq q<3$ & $\bs_2<b\leq b_1^*$ ~~& $2<b<b_1^*$ & Fig.~\ref{fig_bi6cases}(a)\\
& $3\leq q<q_*$ & $\bs_2<b\leq b_1^*$ & $\bs_1<b<b_1^*$ & Fig.~\ref{fig_bi6cases}(d)\\
& $q_*\leq q\leq 27$ & $\bs_2<b<\bs_1$ & n/a & Fig.~\ref{fig_bi6cases}(c)
\\ \hline
(ii) $3\leq r_*<\tilde{r}$ & $0\leq q<3$ & $\bs_2<b\leq b_1^*$ & $2<b<b_1^*$ & Fig.~\ref{fig_bi6cases}(a)\\
& $3\leq q<q_*$ & $\bs_2<b\leq b_1^*$ & $\bs_1<b<b_1^*$ & Fig.~\ref{fig_bi6cases}(d)\\
& $q_*\leq q<27$ & $b_2^*\leq b\leq b_1^*$ & $b_2^*<b<\bs_2$, $\bs_1<b<b_1^*$ ~~& Fig.~\ref{fig_bi6cases}(e)\\
& $27\leq q\leq \qmax$ ~~& $b_2^*\leq b\leq b_1^*$ & $b_2^*<b<b_1^*$ & Fig.~\ref{fig_bi6cases}(f)
\\ \hline
(iii) $\tilde{r}\leq r_*<4$ & $0\leq q<q_*$ & $\bs_2<b\leq b_1^*$ & $2<b<b_1^*$ & Fig.~\ref{fig_bi6cases}(a)\\
& $q_*\leq q<3$ & $b_2^*\leq b\leq b_1^*$ & $b_2^*<b<\bs_2$, $2<b<b_1^*$ & Fig.~\ref{fig_bi6cases}(b)\\
& $3\leq q<27$ & $b_2^*\leq b\leq b_1^*$ & $b_2^*<b<\bs_2$, $\bs_1<b<b_1^*$ & Fig.~\ref{fig_bi6cases}(e)\\
& $27\leq q\leq \qmax$ & $b_2^*\leq b\leq b_1^*$ & $b_2^*<b<b_1^*$ & Fig.~\ref{fig_bi6cases}(f)
\\ \hline
(iv) $r_*\geq4$ & $0\leq q<3$ & $b_2^*\leq b\leq b_1^*$ & $b_2^*<b<\bs_2$, $2<b<b_1^*$ & Fig.~\ref{fig_bi6cases}(b)\\
& $3\leq q<27$ & $b_2^*\leq b\leq b_1^*$ & $b_2^*<b<\bs_2$, $\bs_1<b<b_1^*$ & Fig.~\ref{fig_bi6cases}(e)\\
& $27\leq q\leq \qmax$ & $b_2^*\leq b\leq b_1^*$ & $b_2^*<b<b_1^*$ & Fig.~\ref{fig_bi6cases}(f)
\\ \hline\hline
\end{tabular} 
\label{table_necessary}
\end{table*}
\begin{figure}[t]
\centering
\includegraphics[width=8cm]{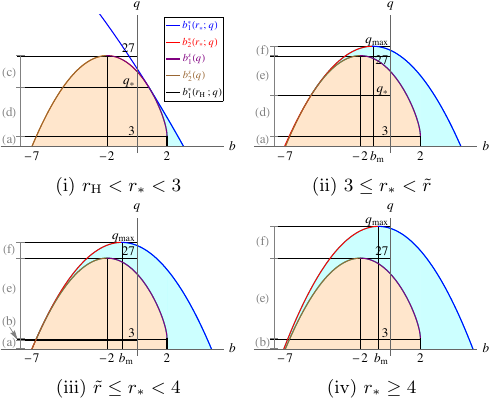}
\caption{Typical parameter region in the $b$-$q$ plane satisfying the necessary condition for photon escape from $r=r_*$ to infinity. The blue, red, purple, and brown curves denote $b=b_1^*(r_*;q)$, $b_2^*(r_*;q)$, $\bs_1(q)$, and $\bs_2(q)$, respectively. The black segment denotes $b=b_1(\rh;q)=2$ and $q\in[0,3]$. The orange regions represent the parameter regions where only photons initially emitted outward (i.e., $\s_r=+$) satisfy the necessary condition for escape. The cyan regions represent the parameter regions where photons initially emitted both outward and inward (i.e., $\s_r=\pm$) satisfy the necessary condition for escape. The labels (a)--(f) in the figure represent the part labels in Fig.~\ref{fig_bi6cases}.}
\label{fig_necessary}
\end{figure}
Let us summarize the necessary condition for photon escape in the ($b,q$) parameter region for a fixed $r_*$. To perform it, we introduce four ranges of $r_*$:
\begin{align}
(\mathrm{i})&~\rh<r_*<3,\\
(\mathrm{ii})&~3\leq r_*<\tilde{r},\\
(\mathrm{iii})&~\tilde{r}\leq r_*<4,\\
(\mathrm{iv})&~r_*\geq4,
\end{align}
where $\tilde{r}\=r_2(3)= 3.95137...$ is the largest solution of $\qSPO(r)=3$. In addition, we introduce two specific values of $q$,
\begin{align}
q_*(r_*)&\=\qSPO(r_*),\\
\qmax(r_*)&\=\frac{r_*^3}{r_*-2}.
\end{align}
When $q=\qmax$, the functions $b_i^*$ coincide with each other, and their value is  
\begin{align}
\bm\= b_i(r_*;\qmax)=\frac{2}{2-r_*}.
\end{align}
When $q>\qmax$, the position $r_*$ enters the forbidden region. Therefore, the maximum value of $q$ is limited by $\qmax$. The necessary conditions for photon escape are summarized in Table~\ref{table_necessary}.%
\footnote{Note that in Refs.~\cite{Ogasawara:2019mir,Ogasawara:2020frt}, the photon escape regions were considered only for the case (i), i.e., $r_{\mathrm{H}}<r_*<3$. Here, we identify them for the entire range of $r_*$.}
It is useful to visualize the necessary condition for photon escape in the $b$-$q$ plane; see Fig.~\ref{fig_necessary}. The blue, red, purple, and brown curves denote $b=b_1^*(r_*;q)$, $b_2^*(r_*;q)$, $\bs_1(q)$, and $\bs_2(q)$, respectively. The black segment denotes $b=b_1(\rh;q)=2$ and $q\in[0,3]$. The orange regions show the parameter regions where photons initially emitted outward (i.e., $\s_r=+$) satisfy the necessary condition for escape. The cyan regions show the parameter regions where photons initially emitted both outward and inward (i.e., $\s_r=\pm$) satisfy the necessary condition for escape.
Figures \ref{fig_necessary}(i)--\ref{fig_necessary}(iv) correspond to Tables \ref{table_necessary}(i)--\ref{table_necessary}(iv), respectively.

\subsection{Necessary and sufficient condition for photon escape, $q\geq0$}
Let us further restrict the above necessary condition for photon escape by the non-negativity of $\Theta$, i.e., Eq.~\eqref{allowed_bTheta}. The common region of these conditions provides the necessary and sufficient parameter region in which a photon can escape from $(r,\theta)=(r_*,\theta_*)$ to infinity. We call it the escapable region. An example of the escapable region is seen in Fig.~\ref{fig_egER}. The green curve denotes $\Theta=0$, and the other curves and colored regions are the same as in Fig.~\ref{fig_necessary}. We can find that the escapable region corresponds to the regions in Fig.~\ref{fig_necessary} further restricted by the condition \eqref{allowed_bTheta}.

\subsection{Necessary and sufficient condition for photon escape, $q<0$}
\label{subsec:3-3}
\begin{figure}[t]
\centering
\includegraphics[width=8cm]{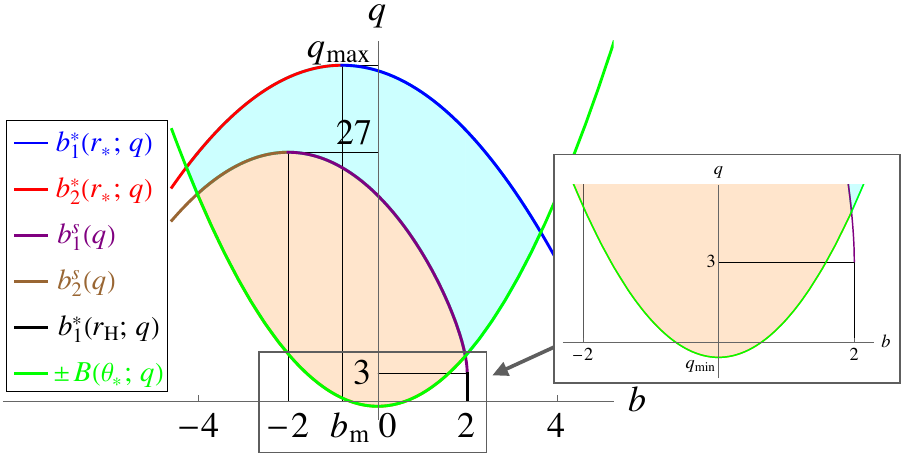}
\caption{Typical shape of the escapable region. The green curve denotes $\Theta=0$, or equivalently, $b=\pm B(\theta_*;q)$. The other curves and colored regions are the same as in Fig.~\ref{fig_necessary}.
}
\label{fig_egER}
\end{figure}

We identify the escapable region for $q<0$. The negative $q$ together with the non-negativity of $\Theta$ at $\theta=\theta_*$ leads to
\begin{align}
\cot^2\theta_*\left(b^2-\sin^2\theta_*\right)\leq q<0.
\label{ineq_negativeq}
\end{align}
This implies that $|b|<1$ for $q<0$, and the minimum value of $q$ is given at $b=0$ as
\begin{align}
\qmin=-\cos^2\theta_*.
\end{align}
For $q<0$, the SPOs are not relevant to photon escape because they do not exist. Therefore, we only focus on $R\geq0$, or equivalently,
\begin{align}
q\leq\frac{r}{(r-1)^2}\left[r^3+\left(1-b^2\right)r+2(1-b)^2\right].
\label{allowed_negativeq}
\end{align}
The right-hand side is positive for all $|b|<1$ and $r>1$. Hence, the allowed region \eqref{allowed_negativeq} contains the entire parameter region \eqref{ineq_negativeq}. 
This means that any radial turning point no longer appears for $q<0$.
Finally, we conclude that photons with $q<0$ can escape to infinity if they are initially emitted outward (i.e., $\s_r=+$) and take the range \eqref{ineq_negativeq}. Figure \ref{fig_egER} shows an example of the escapable region (see the orange region of $q<0$).
\clearpage

\section{Critical values for classifying photon escape}
\label{sec:4}
In order to classify the escapable region completely, we introduce the critical polar angles and the critical values of $q$.

\subsection{Critical angles}
We introduce four critical polar angles $\theta_1$, $\theta_2$, $\theta_3$, and $\tm$, at which the classification of the escapable region varies qualitatively.
Solving $\Theta=0$ for $\theta$, we obtain a solution
\begin{align}
\theta=\tilde{\theta}(b,q)=\arcsin\left[\frac{\left|\sqrt{(b+1)^2+q}-\sqrt{(b-1)^2+q}\right|}{2}
\right];
\end{align}
see Appendix \ref{App:ThetaEq}.

\begin{figure}[t]
\centering
\includegraphics[width=8cm]{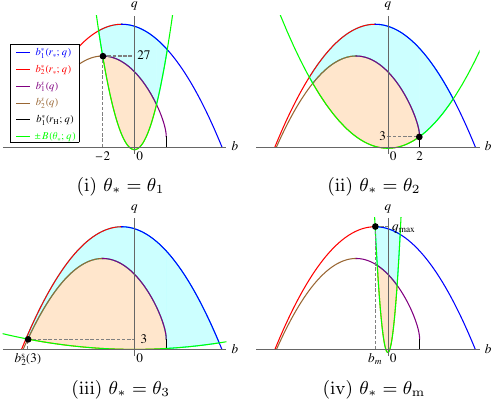}
\caption{Escapable region for $\theta_*$ being the critical angles (i) $\theta_*=\theta_1$, (ii) $\theta_*=\theta_2$, (iii) $\theta_*=\theta_3$, and (iv) $\theta_*=\tm(r_*)$. Four black dots denote $(b,q)=(-2,27)$, $(2,3)$, $(\bs_2(3),3)$, and $(\bm(r_*),\qmax(r_*))$. Each curve and colored region is the same as Fig.~\ref{fig_egER}.}
\label{fig_criticaltheta}
\end{figure}

The first special point is $(b,q)=(-2,27)$, where $\bs_1$ and $\bs_2$ coincide with each other. We define $\theta_1$ as $\theta_*$ at which $-B(\theta_*;q)$ passes through $(-2,27)$, i.e., $B(\theta_1;27)=2$ [see the black dot in Fig.~\ref{fig_criticaltheta}(i)]. Then, $\theta_1$ is given by
\begin{align}
\theta_1=\tilde{\theta}(-2,27)=
\arcsin\left(3-\sqrt{7}\right)\simeq20.7^\circ.
\end{align}
When $\theta_*<\theta_1$, $\bs_2<-B$ holds in the range $q\leq27$. This implies that the minimum value of $b$ in the escapable region for $q\in[\qmin,27]$ is always $-B$.

The second special point is $(b,q)=(2,3)$, where $r_1=\rh=1$ and $\bs_1=b_1(\rh;q)=2$. We define $\theta_2$ as $\theta_*$ at which $B(\theta_*;q)$ passes through $(2,3)$, i.e., $B(\theta_2;3)=2$ [see the black dot in Fig.~\ref{fig_criticaltheta}(ii)]. Then, $\theta_2$ is given by
\begin{align}
\theta_2=\tilde{\theta}(2,3)=
\arcsin\left(\sqrt{3}-1\right)\simeq47.1^\circ.
\end{align}
When $\theta_*<\theta_2$, $B<2$ holds in the range $q\leq3$. This implies that the maximum value of $b$ in the escapable region for $q\in[\qmin,3]$ is always $B$.

The third special point is $(b,q)=(\bs_2(3),3)$, where $\bs_2(3)\simeq -6.71$. We define $\theta_3$ as $\theta_*$ at which $-B(\theta_*;q)$ passes through $(\bs_2(3),3)$, i.e., $-B(\theta_3;3)=\bs_2(3)$ [see the black dot in Fig.~\ref{fig_criticaltheta}(iii)]. Then, $\theta_3$ is given by
\begin{align}
\theta_3=\tilde{\theta}(\bs_2(3)),3)
\simeq 75.4^\circ.
\end{align}
When $\theta_*<\theta_3$, $\bs_2<-B$ holds in the range $q\leq3$. This implies that the minimum value of $b$ in the escapable region for $q\in[\qmin,3]$ is always $-B$.

The fourth special point is $(b,q)=(\bm,\qmax)$, where $b_1^*$ and $b_2^*$ coincide with each other. We define $\tm$ as $\theta_*$ at which $-B(\theta_*;q)$ passes through $(\bm,\qmax)$, i.e., $-B(\tm;\qmax)=\bm(r_*)$ [see the black dot in Fig.~\ref{fig_criticaltheta}(iv)]. Then, $\tm$ is given by
\begin{align}
\tm(r_*)&=\tilde{\theta}(\bm(r_*),\qmax(r_*))
\nonumber\\
&=\arcsin\left[\frac{r_*(r_*-1)-\sqrt{r^2_*(r_*-1)^2-8(r_*-2)}}{2(r_*-2)}
\right].
\end{align}
Note that we only need to consider $\tm$ for $r_*\geq3$ because it does not contribute to specifying the escapable region when $r_*<3$. The critical angle $\tm$ depends on $r_*$ and monotonically decreases with $r_*$ in the range
\begin{align}
\tm(\infty)=0<\tm(r_*)\leq\theta_1=\tm(3).
\end{align}
When $\theta_*<\tm$, $b_2^*<-B$ holds in the range $q\leq\qmax$. This implies that the minimum value of $b$ in the escapable region for $q\in[\qmin,\qmax]$ is always $-B$.

\subsection{Critical values of $q$}
\begin{figure*}[t]
\centering
\includegraphics[width=16cm]{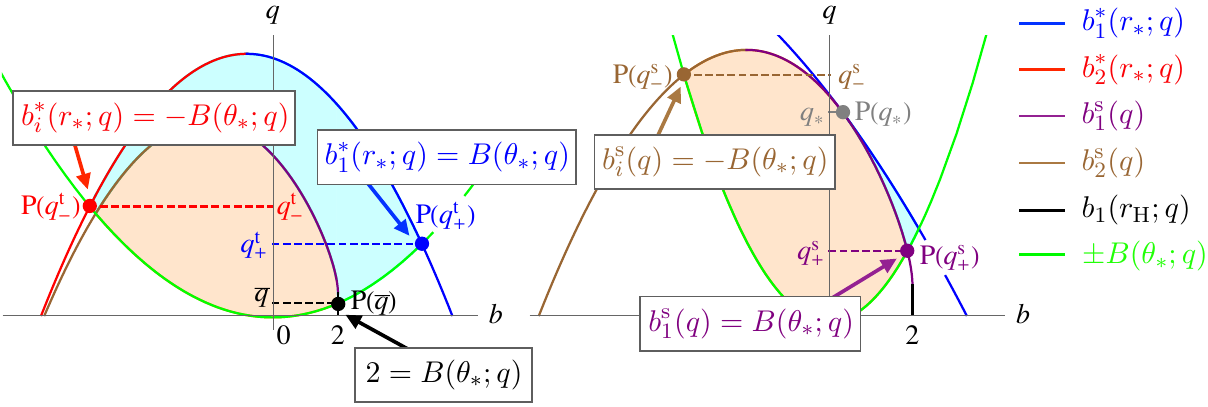}
\caption{Five critical values of $q$. Each curve and colored region is the same as Fig.~\ref{fig_egER}.}
\label{fig_criticalq}
\end{figure*}

We introduce five critical values $\qb$, $\qt_\pm$, and $\qs_\pm$, as the values of $q$ at the intersections of $b=2$, $b_i^*(r_*;q)$, $\bs_i(q)$, and $\pm B(\theta_*;q)$, at which the classification of the parameter ranges varies qualitatively.

We define $\qb$ as the value of $q$ at the intersection of $b=2$ and $b=B(\theta_*;q)$ in the range $0\leq q\leq3$, and we denote the intersection as P$(\qb)$ (see the black dot in Fig.~\ref{fig_criticalq}).
Then, $\qb$ is given by%
\footnote{This $\qb(\theta_*)$ is expressed as $q_1(\theta_*)$ in Ref.~\cite{Ogasawara:2020frt}}
\begin{align}
\qb(\theta_*)=\frac{3+\cos^2\theta_*}{\tan^2\theta_*},
\end{align}
which only appears for $\theta_*\in[\theta_2,\pi/2)$ and monotonically decreases with $\theta_*$ in the range
\begin{align}
\qb(\pi/2)=0<\qb(\theta_*)\leq 3=\qb(\theta_2).
\end{align}
When $q<\qb$, $B<b_1^*$ holds. This implies that the maximum value of $b$ in the escapable region for $q<\qb$ is always $B$.

We define $\qt_+$ as the value of $q$ at the intersection of $b=b_1^*(r_*; q)$ and $b=B(\theta_*; q)$, and we denote the intersection as P$(\qt_+)$ (see the blue dot in Fig.~\ref{fig_criticalq}).
On the other hand, we define $\qt_-$ as the value of $q$ at the intersection of $b=b_2^*(r_*; q)$ and $b=-B(\theta_*; q)$ for $\theta_*\geq\tm$ and $b=b_1^*(r_*; q)$ and $b=-B(\theta_*; q)$ for $\theta_*\leq\tm$, and we denote the intersection as P$(\qt_-)$ (see the red dot in Fig.~\ref{fig_criticalq}).
Then, $\qt_\pm$ are given by%
\footnote{These $\qt_+(r_*,\theta_*)$ and $\qt_-(r_*,\theta_*)$ are expressed as $q_2(r_*,\theta_*)$ and $q_6(r_*,\theta_*)$, respectively, in Ref.~\cite{Ogasawara:2020frt}.}%
\begin{align}
&\qt_\pm(r_*,\theta_*) \nonumber\\
&=\cos^2\theta_*\left[\left(\frac{(r_*-1)\left(r_*^2+\cos^2\theta_*\right) \mp 2r_*\sin\theta_*}{(r_*-1)^2-\sin^2\theta_*}
\right)^2-1\right],
\end{align}
where $\qt_-$ only appears for $r_*>2$. For $\theta_*=\tm$, $b=b_i^*$ and $b=-B$ coincide with each other at $q=\qmax$. Note that $\qt_+<\qt_-$ holds for all $r_*$ and $\theta_*$. Figures \ref{fig_qstarqtPM}(i) and \ref{fig_qstarqtPM}(ii) show the values of $\qt_+$ and $\qt_-$ in the $r_*$-$\theta_*$ parameter space, respectively.

We define $\qs_+$ as the value of $q$ at the intersection of $b=\bs_1(q)$ and $b=B(\theta_*;q)$, and we denote the intersection as P$(\qs_+)$ (see the purple dot in Fig.~\ref{fig_criticalq}).
On the other hand, we define $\qs_-$ as the value of $q$ at the intersection of $b=\bs_2(q)$ and $b=-B(\theta_*;q)$ for $\theta_*\geq\theta_1$ and $b=\bs_1(q)$ and $b=-B(\theta_*;q)$ for $\theta_*\leq\theta_1$, and we denote the intersection as P$(\qs_-)$ (see the brown dot in Fig.~\ref{fig_criticalq}).
Then, $\qs_\pm$ are given by%
\footnote{This $\qs_+(\theta_*)$ is expressed as $q_3(\theta_*)$, and $\qs_-(\theta_*)$ is expressed as $q_4(\theta_*)$ for $\theta_*\geq\theta_1$ and as $q_5(\theta_*)$ for $\theta_*<\theta_1$, respectively, in Ref.~\cite{Ogasawara:2020frt}.}%
\begin{align}
\qs_\pm(\theta_*)&=\qSPO(x_\pm)=-(x_\pm)^4+4x_\pm, \\
x_\pm&\=1\mp\sin\theta_*+\sqrt{2(1\mp\sin\theta_*)},
\end{align}
where $\qs_+$ only appears for $\theta_*\in(0,\theta_2)$ and monotonically decreases with $\theta_*$ in the range
\begin{align}
\qs_+(\theta_2)=3<\qs_+(\theta_*)<11+8\sqrt{2}=\qs_+(0).
\end{align}
As $\theta_*$ increases from $0$ to $\pi/2$, $\qs_-$ begins with $\qs_-(0)=11+8\sqrt{2}$, monotonically increases to the maximum value $27$ at $\theta_*=\theta_1$, and monotonically decreases to $\qs_-(\pi/2)=0$.
For $\theta_*=\theta_1$, $b=\bs_i$ and $b=-B$ coincide with each other at $q=27$.

In addition, we define $\mathrm{P}(q_*)$ as a point in the $b$-$q$ plane that represents $(b_1^*(r_*;q_*),q_*)$ for $r_*<3$ and $(b_2^*(r_*;q_*),q_*)$ for $3\leq r_*\leq4$ (see the gray dot in Fig.~\ref{fig_criticalq}).

These are summarized in Table \ref{table_5points}.

\begin{table}[t]
\centering
\caption{Definition of special points P on the $b$-$q$ plane and the values of $q$ and $b$ at these points.}
\begin{tabular}{llll}
\hline\hline
P & Intersection & $q$ & $b$
\\ \hline
$\mathrm{P}(\qb)$ & $b=2$ and $b=B$ & $\qb(\theta_*)$ & $2=B(\theta_*;\qb)$\\
$\mathrm{P}(\qt_+)$ & $b=b_1^*$ and $b=B$ & $\qt_+(r_*,\theta_*)$ & $b_1^*(r_*;\qt_+)=B(\theta_*;\qt_+)$\\
$\mathrm{P}(\qt_-)$ & $b=b_i^*$ and $b=-B$ & $\qt_-(r_*,\theta_*)$ & $b_i^*(r_*;\qt_-)=-B(\theta_*;\qt_-)$\\
$\mathrm{P}(\qs_+)$ & $b=\bs_1$ and $b=B$ & $\qs_+(\theta_*)$ & $\bs_1(\qs_+)=B(\theta_*;\qs_+)$\\
$\mathrm{P}(\qs_-)$ & $b=\bs_i$ and $b=-B$ & $\qs_-(\theta_*)$ & $\bs_i(\qs_-)=-B(\theta_*;\qs_-)$\\
$\mathrm{P}(q_*)$ & -- & $q_*(r_*)$ & $b_1^*(r_*;q_*)$ for $r_*<3$\\
&&& ~$b_2^*(r_*;q_*)$ for $3\leq r_*\leq 4$
\\ \hline\hline
\end{tabular} 
\label{table_5points}
\end{table}

\subsection{Conditions for critical values of $q$ involved in classification}
\begin{figure}[t]
\centering
\includegraphics[width=8.5cm]{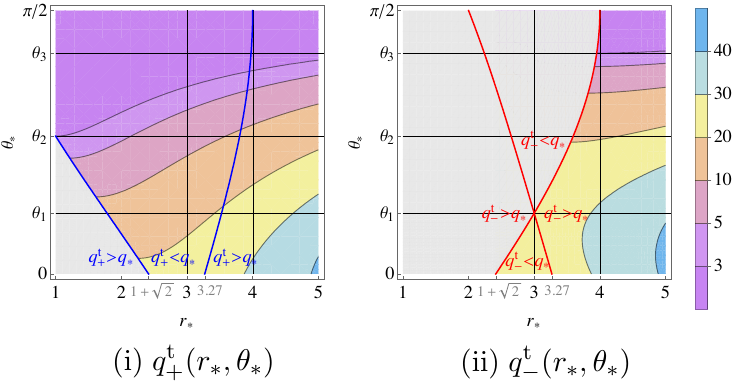}
\caption{Values of $\qt_\pm(r_*,\theta_*)$ in the $r_*$-$\theta_*$ parameter space. Blue and red curves denote $\qt_+=q_*$ and $\qt_-=q_*$, respectively. Gray regions in (i) and (ii) represent the parameter regions where the points $\mathrm{P}(\qt_+)$ and $\mathrm{P}(\qt_-)$ do not contribute to specifying the escapable region, respectively.
}
\label{fig_qstarqtPM}
\end{figure}
\begin{figure}[b]
\centering
\includegraphics[width=8cm]{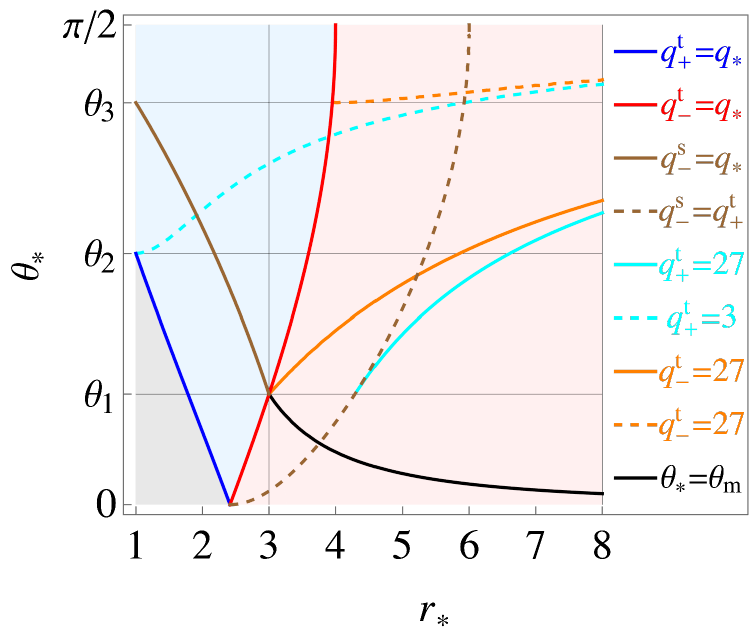}
\caption{Relationship of the characteristic $q$'s in the $r_*$-$\theta_*$ parameter space. In the gray region, $\qt_\pm$ and $q_*$ do not contribute to specifying the escapable region. In the blue region, $\qt_+$ and $q_*$ contribute to specifying the escapable region, while $\qt_-$ does not. In the red region, $\qt_\pm$ contribute to specifying the escapable region, while $q_*$ does not. Note that curves representing $\qt_+=q_*$ for $r_*\geq3$ and $\qt_-=q_*$ for $r_*<3$ and $\theta_*\geq\theta_1$, and for $r_*\geq3$ and $\theta_*<\theta_1$ are not plotted. This is because these curves do not contribute to the classification of photon escape.
}
\label{fig_qall}
\end{figure}

It is worth noting that $\qt_\pm$ and $q_*$ are not always involved in classification; i.e., the intersections P($\qt_+$), P($\qt_-$), and P$(q_*)$ are not always involved in classification.

When $\qt_+>q_*$ for $r_*<3$, the intersection P($\qt_+$) does not contribute to specifying the escapable region [see the gray region in Fig.~\ref{fig_qstarqtPM}(i)]. On the other hand, when $\qt_+\leq q_*$ for $r_*<3$, the intersection P($\qt_+$) is a special point, which contributes to specifying the escapable region. For $r_*\geq3$, the intersection P($\qt_+$) always contributes to it. The colored region in Fig.~\ref{fig_qstarqtPM}(i) represents the parameter region where $\qt_+$ contributes to specifying the escapable region.

For the same reason as for $\qt_+$, the relative values of $\qt_-$ and $q_*$ determine whether $\qt_-$ contributes to specifying the escapable region.
In the case of $r_*<3$, only when $\qt_-\leq q_*$ for $\theta_*<\theta_1$, the intersection P($\qt_-$) is included in the escapable region. In the case of $r_*\geq3$, when $\qt_-\leq q_*$ for $\theta_*<\theta_1$ and when $\qt_-\geq q_*$, the intersection P$(\qt_-)$ is included. The colored region in Fig.~\ref{fig_qstarqtPM}(ii) represents the parameter region where $\qt_-$ contributes to specifying the escapable region.

In the case of $r_*<3$, when $\qt_+\leq q_*\leq\qt_-$ and when $\qt_-\leq q_*$ for $\theta_*\geq\theta_1$, the point P($q_*$) contributes to specifying the escapable region. In the case of $r_*\geq3$, the point P($q_*$) contributes to it only when $\qt_-\leq q_*$ for $\theta_*\geq\theta_1$. The blue region in Fig.~\ref{fig_qall} represents the parameter region where $q_*$ contributes to specifying the escapable region.

Other intersections P($\qb$) and P($\qs_+$) contribute to specifying the escapable region when $\theta_*\geq\theta_2$ and $\theta_*<\theta_2$, respectively, and P($\qs_-$) always contributes to it.

In the following section, we will perform a complete classification of photon escape.

\section{Complete classification of photon escape in the extremal Kerr black hole}
\label{sec:5}
\begin{table}[t]
\centering
\caption{Definition of each class and the characteristic values of $q$ that appear in the classification in each class.
}
\begin{tabular}{lll}
\hline\hline
Class & Range of $(r_*,\theta_*)$ & Characteristic $q$'s
\\ \hline
I & $r_*<3$ and $\theta_*\in(0,\theta_1)$ & $q_*$, $\qt_\pm$, and $\qs_\pm$\\
II & $r_*<3$ and $\theta_*\in[\theta_1,\theta_2)$ & $q_*$, $\qt_+$, and $\qs_\pm$\\
III & $r_*<3$ and $\theta_*\in[\theta_2,\theta_3)$ & $q_*$, $\qb$, $\qt_+$, $\qs_-$, and $3$\\
IV & $r_*<3$ and $\theta_*\in[\theta_3,\pi/2)$ & $q_*$, $\qb$, $\qt_+$, $\qs_-$, and $3$\\
V & $r_*\geq3$ and $\theta_*\in(0,\theta_1)$ & $\qt_\pm$ and $\qs_\pm$\\
VI & $r_*\geq3$ and $\theta_*\in[\theta_1,\theta_2)$ & $q_*$, $\qt_\pm$, $\qs_\pm$, and $27$\\
VII & $r_*\geq3$ and $\theta_*\in[\theta_2,\theta_3)$ & $q_*$, $\qb$, $\qt_\pm$, $\qs_-$, $3$, and $27$\\
VIII & $r_*\geq3$ and $\theta_*\in[\theta_3,\pi/2)$ & $q_*$, $\qb$, $\qt_\pm$, $\qs_-$, $3$, and $27$
\\ \hline\hline
\end{tabular} 
\label{table_8class}
\end{table}

In this section, we make a complete classification of photon escape. We define eight classes according to a spacetime position of an emitter $(r_*,\theta_*)$ (see Table \ref{table_8class}):
\begin{align}
\mathrm{Class~I}&: r_*<3 ~\mathrm{and}~ 0<\theta_*<\theta_1,\\
\mathrm{Class~II}&: r_*<3 ~\mathrm{and}~ \theta_1\leq\theta_*<\theta_2,\\
\mathrm{Class~III}&: r_*<3 ~\mathrm{and}~ \theta_2\leq\theta_*<\theta_3,\\
\mathrm{Class~IV}&: r_*<3 ~\mathrm{and}~ \theta_3\leq\theta_*<\pi/2,\\
\mathrm{Class~V}&: r_*\geq 3 ~\mathrm{and}~ 0<\theta_*<\theta_1,\\
\mathrm{Class~VI}&: r_*\geq3 ~\mathrm{and}~ \theta_1\leq\theta_*<\theta_2,\\
\mathrm{Class~VII}&: r_*\geq3 ~\mathrm{and}~ \theta_2\leq\theta_*<\theta_3,\\
\mathrm{Class~VIII}&: r_*\geq3 ~\mathrm{and}~ \theta_3\leq\theta_*<\pi/2.
\end{align}
For $r_*<3$ (i.e., classes I--IV), $q_*$ monotonically increases with $r_*$ in the range
\begin{align}
\qSPO(1)=3<q_*(r_*)<27=\qSPO(3),
\end{align}
and we only need to consider the range of $\qmin\leq q\leq27$ because there is no escapable region in $q>27$.
For $r_*\geq3$ (i.e., classes V--VIII), as $r_*$ increases from $3$ to $\infty$, $q_*$ begins at $\qSPO(3)=27$ and monotonically decreases to $-\infty$. We only need to consider the range of $\qmin\leq q\leq\qmax$ because there is no escapable region in $q>\qmax$.

Figure \ref{fig_qall} shows the relationship of the characteristic $q$'s in the $r_*$-$\theta_*$ parameter space. In the gray region, $\qt_\pm$ and $q_*$ do not contribute to specifying the escapable region. In the blue region, $\qt_+$ and $q_*$ contribute to specifying the escapable region, while $\qt_-$ does not. In the red region, $\qt_\pm$ contribute to specifying the escapable region, while $q_*$ does not. 
For each class, the regions separated by these curves give different escapable parameter regions. For example, since the region $r_*<3$ and $0<\theta_*<\theta_1$ (i.e., class I) is divided into four by three curves, there are four different cases of the escapable regions.
Note that the curve representing $\qt_+=q_*$ for $r_*\geq3$ is not plotted because it does not contribute to the classification of photon escape. Also, for the same reason, the curve representing $\qt_-=q_*$ for $r_*<3$ and $\theta_*\geq\theta_1$, and for $r_*\geq3$ and $\theta_*<\theta_1$ is not plotted. 

In the following subsections, we consider the escapable region separately for each class.

\subsection{Class I: $r_*<3$ and $0<\theta_*<\theta_1$}
\begin{figure}[b]
\centering
\includegraphics[width=6cm]{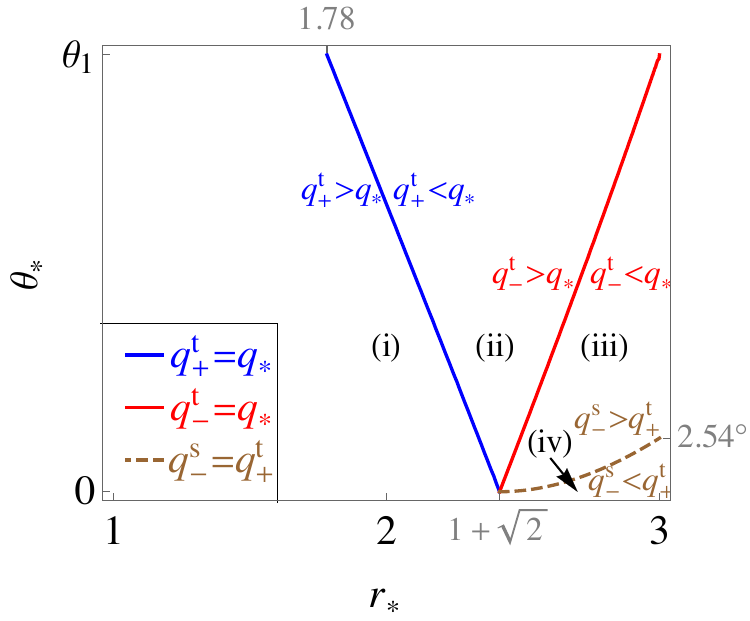}
\caption{(Class I) Relationship of the characteristic $q$'s.}
\label{fig_q1}
\end{figure}
\begin{figure}[H]
\centering
\includegraphics[width=8cm]{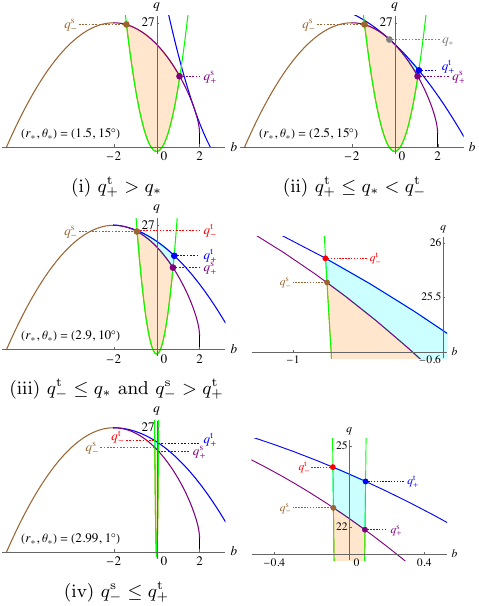}
\caption{(Class I) Typical shape of the escapable region for $r_*<3$ and $0<\theta_*<\theta_1$. The blue, purple, brown, and green curves denote $b=b_1^*(r_*;q)$, $\bs_1(q)$, $\bs_2(q)$, and $\pm B(\theta_*;q)$, respectively. The gray, blue, red, purple, and brown dashed lines denote $q=q_*(r_*)$, $\qt_+(r_*,\theta_*)$, $\qt_-(r_*,\theta_*)$, $\qs_+(\theta_*)$, and $\qs_-(\theta_*)$, respectively.}
\label{fig_ER1}
\end{figure}

There exist four cases according to the relative values of $q_*$, $\qt_\pm$, and $\qs_-$ (see Fig.~\ref{fig_q1}):
\begin{align}
(\mathrm{i})&~ \qt_+>q_*,\\
(\mathrm{ii})&~ \qt_+\leq q_*<\qt_-, \label{q_Class_I_ii}\\
(\mathrm{iii})&~ \qt_-\leq q_* ~\mathrm{and}~ \qs_->\qt_+, \label{q_Class_I_iii}\\
(\mathrm{iv})&~ \qs_-\leq\qt_+,
\end{align}
where case (iv) appears only when $\theta_*<2.543^\circ$.
The equal signs of Eqs.~\eqref{q_Class_I_ii} and \eqref{q_Class_I_iii} hold only when $\qs_+=\qt_+=q_*$ and $\qs_-=\qt_-=q_*$, respectively.
Note that $q_*$ contributes to specifying the escapable region only for case (ii), $\qt_+$ contributes to that for cases (ii)--(iv), and $\qt_-$ contributes to that for cases (iii) and (iv).
The escapable regions in the above cases are summarized in Table~\ref{table_ER1} and Fig.~\ref{fig_ER1}.

\newpage

\begin{table}[H]
\centering
\caption{(Class I) Escapable region $(b,q)$ for $r_*<3$ and $0<\theta_*<\theta_1$.
}
\begin{tabular}{llll}
\hline\hline
Case & $q$ & $b$ ($\s_r=+$) & $b$ ($\s_r=-$)
\\ \hline
(i)--(iv) & $\qmin\leq q<\qs_+$ & $-B\leq b\leq B$ & n/a \\ \hline
(ii) and (iii) & $\qs_+\leq q<\qt_+$ & $-B\leq b\leq B$ & $\bs_1<b\leq B$ \\
(iv) & $\qs_+\leq q<\qs_-$ && \\ \hline
(ii)& $\qt_+\leq q<q_*$ & $-B\leq b\leq b_1^*$ & $\bs_1<b<b_1^*$ \\
(iii)& $\qt_+\leq q<\qs_-$ && \\ \hline
(i) & $\qs_+\leq q<\qs_-$ & $-B\leq b<\bs_1$ & n/a \\
(ii) & $q_*\leq q<\qs_-$ && \\ \hline
(iv)& $\qs_-\leq q<\qt_+$ & $-B\leq b\leq B$ & $-B\leq b\leq B$ \\ \hline
(iii)& $\qs_-\leq q<\qt_-$ & $-B\leq b\leq b_1^*$ & $-B\leq b<b_1^*$ \\
(iv)& $\qt_+\leq q<\qt_-$ && \\ \hline
(i) and (ii)& $\qs_-\leq q\leq27$ & n/a & n/a \\
(iii) and (iv)& $\qt_-\leq q\leq 27$ & n/a & n/a
\\ \hline\hline
\end{tabular} 
\label{table_ER1}
\end{table}

\newpage

\subsection{Class II: $r_*<3$ and $\theta_1\leq\theta_*<\theta_2$}
\begin{figure}[t]
\centering
\includegraphics[width=6cm]{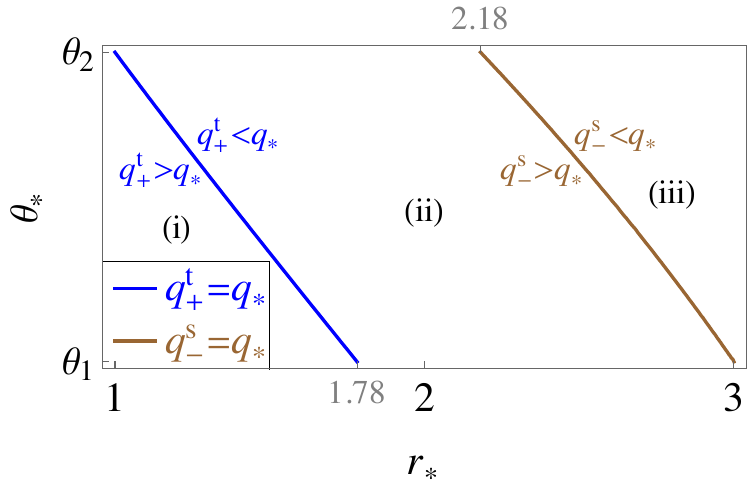}
\caption{(Class II) Relationship of the characteristic $q$'s.}
\label{fig_q2}
\end{figure}
\begin{figure}[H]
\centering
\includegraphics[width=8cm]{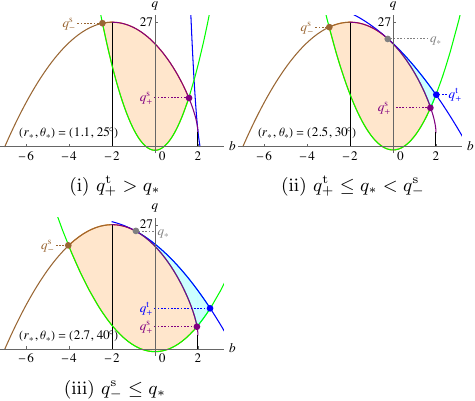}
\caption{(Class II) Typical shape of the escapable region for $r_*<3$ and $\theta_1\leq\theta_*<\theta_2$. The blue, purple, brown, and green curves denote $b=b_1^*(r_*;q)$, $\bs_1(q)$, $\bs_2(q)$, and $\pm B(\theta_*;q)$, respectively. The gray, blue, purple, and brown dashed lines denote $q=q_*(r_*)$, $\qt_+(r_*,\theta_*)$, $\qs_+(\theta_*)$, and $\qs_-(\theta_*)$, respectively.}
\label{fig_ER2}
\end{figure}

There exist three cases according to the relative values of $q_*$, $\qt_+$, and $\qs_-$ (see Fig.~\ref{fig_q2}):
\begin{align}
(\mathrm{i})&~ \qt_+>q_*,\\
(\mathrm{ii})&~ \qt_+\leq q_*<\qs_-, \label{q_Class_II_ii}\\
(\mathrm{iii})&~\qs_-\leq q_*,
\end{align}
where the equal sign of Eq.~\eqref{q_Class_II_ii} holds only when $\qs_+=\qt_+=q_*$. For case (i), $q_*$ and $\qt_+$ do not contribute to specifying the escapable region. The escapable regions in the above cases are summarized in Table~\ref{table_ER2} and Fig.~\ref{fig_ER2}.

\subsection{Class III: $r_*<3$ and $\theta_2\leq\theta_*<\theta_3$}
\begin{figure}[H]
\centering
\includegraphics[width=7cm]{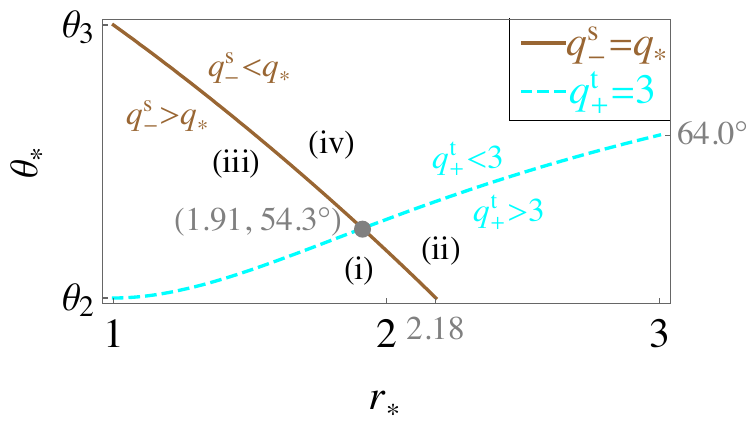}
\caption{(Class III) Relationship of the characteristic $q$'s.}
\label{fig_q3}
\end{figure}

\begin{figure}[H]
\centering
\includegraphics[width=8cm]{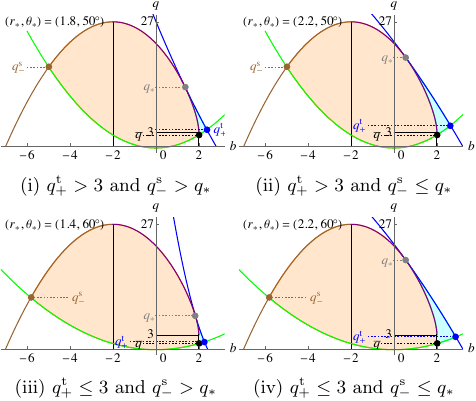}
\caption{(Class III) Typical shape of the escapable region for $r_*<3$ and $\theta_2\leq\theta_*<\theta_3$. The blue, purple, brown, and green curves denote $b=b_1^*(r_*;q)$, $\bs_1(q)$, $\bs_2(q)$, and $\pm B(\theta_*;q)$, respectively. The gray, black, blue, and brown dashed lines denote $q=q_*(r_*)$, $\qb(\theta_*)$, $\qt_+(r_*,\theta_*)$, and $\qs_-(\theta_*)$, respectively.}
\label{fig_ER3}
\end{figure}

There are four cases according to the relative values of $\qt_+$ and $3$, and $\qs_-$ and $q_*$ (see Fig.~\ref{fig_q3}):
\begin{align}
(\mathrm{i})&~ \qt_+>3 ~\mathrm{and}~ \qs_->q_*,\\
(\mathrm{ii})&~ \qt_+>3 ~\mathrm{and}~ \qs_-\leq q_*,\\
(\mathrm{iii})&~ \qt_+\leq3 ~\mathrm{and}~ \qs_->q_*,\\
(\mathrm{iv})&~ \qt_+\leq3 ~\mathrm{and}~ \qs_-\leq q_*.
\end{align}
The escapable regions in the above cases are summarized in Table~\ref{table_ER3} and Fig.~\ref{fig_ER3}.

\subsection{Class IV: $r_*<3$ and $\theta_3\leq\theta_*<\pi/2$}
\begin{figure}[H]
\centering
\includegraphics[width=8cm]{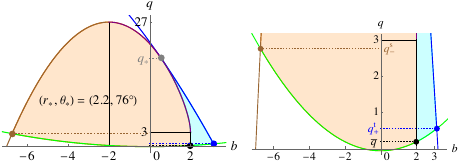}
\caption{(Class IV) Typical shape of the escapable region for $r_*<3$ and $\theta_3\leq\theta_*<\pi/2$. The blue, purple, brown, and green curves denote $b=b_1^*(r_*;q)$, $\bs_1(q)$, $\bs_2(q)$, and $\pm B(\theta_*;q)$, respectively. The gray, black, blue, and brown dashed lines denote $q=q_*(r_*)$, $\qb(\theta_*)$, $\qt_+(r_*,\theta_*)$, and $\qs_-(\theta_*)$, respectively.}
\label{fig_ER4}
\end{figure}

In this case, there is no case classification according to the relative values of $q$. The escapable region in this class is summarized in Table~\ref{table_ER4} and Fig.~\ref{fig_ER4}.

\subsection{Class V: $r_*\geq3$ and $0<\theta_*<\theta_1$}
\begin{figure}[H]
\centering
\includegraphics[width=6cm]{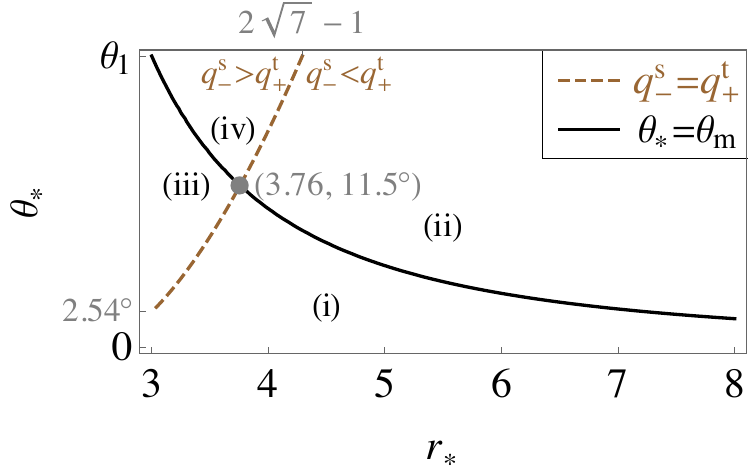}
\caption{(Class V) Relationship between the critical $q$'s and $\tm(r_*)$.}
\label{fig_q5}
\end{figure}

\begin{figure}[H]
\centering
\includegraphics[width=8cm]{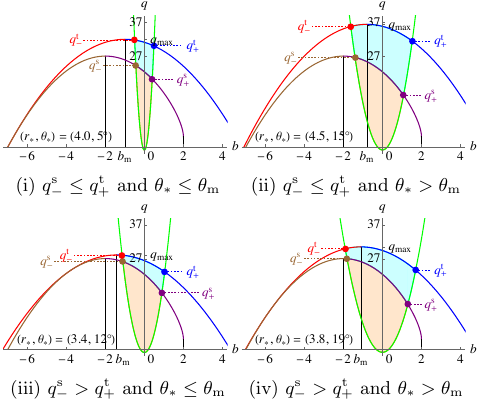}
\caption{(Class V) Typical shape of the escapable region for $r_*\geq3$ and $0<\theta_*<\theta_1$. The blue, purple, brown, and green curves denote $b=b_1^*(r_*;q)$, $\bs_1(q)$, $\bs_2(q)$, and $\pm B(\theta_*;q)$, respectively. The blue, red, purple, and brown dashed lines denote $q=\qt_+(r_*,\theta_*)$, $\qt_-(r_*,\theta_*)$, $\qs_+(\theta_*)$, and $\qs_-(\theta_*)$, respectively.}
\label{fig_ER5}
\end{figure}

There are four cases according to the relative values of $\qt_+$ and $\qs_-$, and $\theta_*$ and $\tm$ (see Fig.~\ref{fig_q5}):
\newpage
\begin{table}[H]
\centering
\caption{(Class II) Escapable region $(b,q)$ for $r_*<3$ and $\theta_1\leq\theta_*<\theta_2$.}
\begin{tabular}{llll}
\hline\hline
Case & $q$ & $b$ ($\s_r=+$) & $b$ ($\s_r=-$)
\\ \hline
(i)--(iii) & $\qmin\leq q<\qs_+$ & $-B\leq b\leq B$ & n/a \\ \hline
(ii) and (iii) & $\qs_+\leq q<\qt_+$ & $-B\leq b\leq B$ & $\bs_1<b\leq B$ \\ \hline
(ii)& $\qt_+\leq q<q_*$ & $-B\leq b\leq b_1^*$ & $\bs_1<b<b_1^*$ \\
(iii)& $\qt_+\leq q<\qs_-$ && \\ \hline
(i) & $\qs_+\leq q<\qs_-$ & $-B\leq b<\bs_1$ & n/a \\
(ii)& $q_*\leq q<\qs_-$ && \\ \hline
(iii)& $\qs_-\leq q<q_*$ & $\bs_2<b\leq b_1^*$ & $\bs_1<b<b_1^*$ \\ \hline
(i) and (ii)& $\qs_-\leq q\leq27$ & $\bs_2<b<\bs_1$ & n/a \\
(iii)& $q_*\leq q\leq 27$ &&
\\ \hline\hline
\end{tabular} 
\label{table_ER2}
\end{table}

\begin{align}
(\mathrm{i})&~\qs_-\leq\qt_+ ~\mathrm{and}~ \theta_*\leq\tm,\\
(\mathrm{ii})&~\qs_-\leq\qt_+ ~\mathrm{and}~ \theta_*>\tm,\\
(\mathrm{iii})&~\qs_->\qt_+ ~\mathrm{and}~ \theta_*\leq\tm,\\
(\mathrm{iv})&~\qs_->\qt_+ ~\mathrm{and}~ \theta_*>\tm.
\end{align}
The escapable regions in the above cases are summarized in Table \ref{table_ER5} and Fig.~\ref{fig_ER5}.
Note that the shape of the escapable region of class V(i) is the same as that of class I(iv), and the shape of the escapable region of class V(iii) is the same as that of class I(iii).

\newpage

\begin{table}[H]
\centering
\caption{(Class III) Escapable region $(b,q)$ for $r_*<3$ and $\theta_2\leq\theta_*<\theta_3$.}
\begin{tabular}{llll}
\hline\hline
Case & $q$ & $b$ ($\s_r=+$) & $b$ ($\s_r=-$)
\\ \hline
(i)--(iv) & $\qmin\leq q<\qb$ & $-B\leq b\leq B$ & n/a \\ \hline
(i) and (ii) & $\qb\leq q<3$ & $-B\leq b\leq B$ & $2<b\leq B$ \\
& $3\leq q<\qt_+$ & $-B\leq b\leq B$ & $\bs_1<b\leq B$ \\ \hline
(iii) and (iv) & $\qb\leq q<\qt_+$ & $-B\leq b\leq B$ & $2<b\leq B$ \\
& $\qt_+\leq q<3$ & $-B\leq b\leq b_1^*$ & $2<b<b_1^*$
\\ \hline
(i) & $\qt_+\leq q<q_*$ & $-B\leq b\leq b_1^*$ & $\bs_1<b<b_1^*$ \\ 
(ii) & $\qt_+\leq q<\qs_-$ \\
(iii) & $3\leq q<q_*$ \\
(iv) & $3\leq q<\qs_-$ \\ \hline
(i) and (iii)& $q_*\leq q<\qs_-$ & $-B\leq b<\bs_1$ & n/a \\
& $\qs_-\leq q\leq 27$ & $\bs_2<b<\bs_1$ & n/a \\ \hline
(ii) and (iv)& $\qs_-\leq q<q_*$ & $\bs_2<b\leq b_1^*$ & $\bs_1<b<b_1^*$ \\
& $q_*\leq q\leq 27$ & $\bs_2<b<\bs_1$ & n/a
\\ \hline\hline
\end{tabular} 
\label{table_ER3}
\end{table}
\begin{table}[H]
\centering
\caption{(Class IV) Escapable region $(b,q)$ for $r_*<3$ and $\theta_3\leq\theta_*<\pi/2$.}
\begin{tabular}{lll}
\hline\hline
$q$ & $b$ ($\s_r=+$) & $b$ ($\s_r=-$)
\\ \hline
$\qmin\leq q<\qb$ & $-B\leq b\leq B$ & n/a \\
$\qb\leq q<\qt_+$ & $-B\leq b\leq B$ & $2<b\leq B$ \\
$\qt_+\leq q<\qs_-$ & $-B\leq b\leq b_1^*$ & $2<b<b_1^*$ \\
$\qs_-\leq q<3$ & $\bs_2<b\leq b_1^*$ & $2<b<b_1^*$ \\
$3\leq q<q_*$ & $\bs_2<b\leq b_1^*$ & $\bs_1<b<b_1^*$ \\
$q_*\leq q\leq 27$ & $\bs_2<b<\bs_1$ & n/a
\\ \hline\hline
\end{tabular} 
\label{table_ER4}
\end{table}
~\\[-10mm]
~\\[-10mm]

\subsection{Class VI: $r_*\geq3$ and $\theta_1\leq\theta_*<\theta_2$}
\begin{figure}[H]
\centering
\includegraphics[width=6cm]{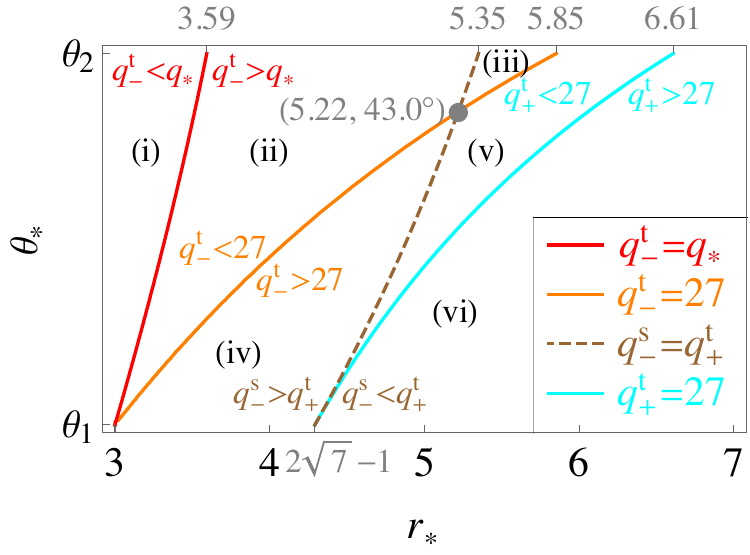}
\caption{(Class VI) Relationship between the critical $q$'s and $q_*$.}
\label{fig_q6}
\end{figure}

\begin{figure}[H]
\centering
\includegraphics[width=8cm]{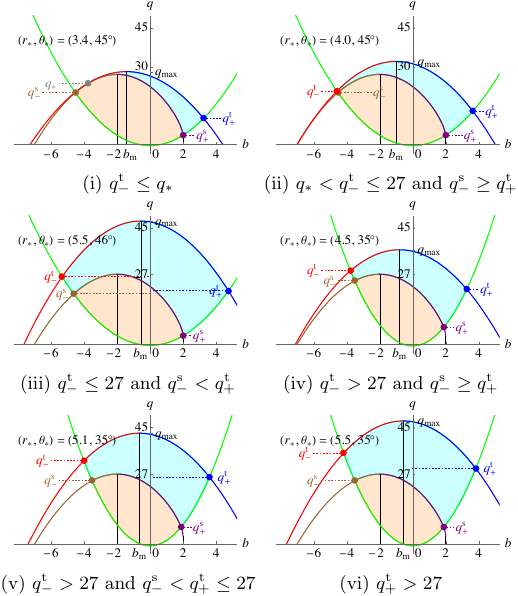}
\caption{(Class VI) Typical shape of the escapable region for $r_*\geq3$ and $\theta_1\leq\theta_*<\theta_2$. The blue, purple, brown, and green curves denote $b=b_1^*(r_*;q)$, $\bs_1(q)$, $\bs_2(q)$, and $\pm B(\theta_*;q)$, respectively. The gray, blue, red, purple, and brown dashed lines denote $q=q_*(r_*)$, $\qt_+(r_*,\theta_*)$, $\qt_-(r_*,\theta_*)$, $\qs_+(\theta_*)$, and $\qs_-(\theta_*)$, respectively.}
\label{fig_ER6}
\end{figure}

There are six cases according to the relative values of $q_*$, $\qt_\pm$, $\qs_-$, and $27$ (see Fig.~\ref{fig_q6}):
\begin{align}
\mathrm{(i)}&~\qt_-\leq q_*,\\
\mathrm{(ii)}&~q_*<\qt_-\leq27 ~\mathrm{and}~ \qs_-\geq\qt_+,\\
\mathrm{(iii)}&~\qt_-\leq27 ~\mathrm{and}~ \qs_-<\qt_+,\\
\mathrm{(iv)}&~\qt_->27 ~\mathrm{and}~ \qs_-\geq\qt_+,\\
\mathrm{(v)}&~\qt_->27 ~\mathrm{and}~ \qs_-<\qt_+\leq27,\\
\mathrm{(vi)}&~\qt_+>27,
\end{align}
where $q_*$ contributes to specifying the escapable region only for case (i), and $\qt_-$ contributes to that for cases (ii)--(vi). The escapable regions in the above cases are summarized in Table~\ref{table_ER6} and Fig.~\ref{fig_ER6}.

\newpage

\begin{table}[H]
\centering
\caption{(Class V) Escapable region ($b,q$) for $r_*\geq3$ and $0<\theta_*<\theta_1$.}
\begin{tabular}{llll}
\hline\hline
Case~~ & $q$ & $b$ ($\s_r=+$) & $b$ ($\s_r=-$) 
\\ \hline
(i)--(iv)& $\qmin\leq q<\qs_+$ & $-B\leq b\leq B$ & n/a\\ \hline
(i) and (ii)&$\qs_+\leq q<\qs_-$ & $-B\leq b\leq B$ & $\bs_1<b\leq B$\\
(iii) and (iv)&$\qs_+\leq q<\qt_+$ && \\ \hline
(i) and (ii)&$\qs_-\leq q<\qt_+$ & $-B\leq b\leq B$ & $-B\leq b\leq B$\\ \hline
(iii) and (iv)&$\qt_+\leq q<\qs_-$ & $-B\leq b\leq b_1^*$ & $\bs_1<b<b_1^*$\\ \hline
(i) and (ii)&$\qt_+\leq q<\qt_-$ & $-B\leq b\leq b_1^*$ & $-B\leq b<b_1^*$\\
(iii) and (iv)&$\qs_-\leq q<\qt_-$ && \\ \hline
(i) and (iii)&$\qt_-\leq q\leq\qmax$ & n/a & n/a\\ \hline
(ii) and (iv)&$\qt_-\leq q\leq\qmax$ & $b_2^*\leq b\leq b_1^*$ & $b_2^*<b<b_1^*$
\\ \hline\hline
\end{tabular}
\label{table_ER5}
\end{table}

\begin{table}[H]
\centering
\caption{(Class VI) Escapable region ($b,q$) for $r_*\geq3$ and $\theta_1\leq\theta_*<\theta_2$.}
\begin{tabular}{llll}
\hline\hline
Case~~ & $q$ & $b$ ($\s_r=+$) & $b$ ($\s_r=-$) 
\\ \hline
(i)--(vi)& $\qmin\leq q<\qs_+$ & $-B\leq b\leq B$ & n/a\\ \hline
(i), (ii), (iv) & $\qs_+\leq q<\qt_+$ & $-B\leq b\leq B$ & $\bs_1<b\leq B$\\
& $\qt_+\leq q<\qs_-$ & $-B\leq b\leq b_1^*$ & $\bs_1<b<b_1^*$\\ \hline
(iii), (v), (vi)& $\qs_+\leq q<\qs_-$ & $-B\leq b\leq B$ & $\bs_1<b\leq B$\\ \hline
(iii) and (v) & $\qs_-\leq q<\qt_+$ & $-B\leq b\leq B$ & $-B\leq b<\bs_2$\\
(vi) & $\qs_-\leq q<27$ && ~and $\bs_1<b\leq B$\\ \hline
(i)& $\qs_-\leq q<q_*$ & $\bs_2<b\leq b_1^*$ & $\bs_1<b<b_1^*$\\ \hline
(ii) & $\qs_-\leq q<\qt_-$ & $-B\leq b\leq b_1^*$ & $-B\leq b<\bs_2$\\
(iii) & $\qt_+\leq q<\qt_-$ && ~and $\bs_1<b<b_1^*$\\
(iv) & $\qs_-\leq q<27$ &&\\ 
(v) & $\qt_+\leq q<27$ && \\ \hline
(vi)& $27\leq q<\qt_+$ & $-B\leq b\leq B$ & $-B\leq b\leq B$\\ \hline
(i)& $q_*\leq q<27$ & $b_2^*\leq b\leq b_1^*$ & $b_2^*<b<\bs_2$\\
(ii) and (iii)& $\qt_-\leq q<27$ && ~and $\bs_1<b<b_1^*$\\ \hline
(iv) and (v)& $27\leq q<\qt_-$ & $-B\leq b\leq b_1^*$ & $-B\leq b<b_1^*$\\ 
(vi)& $\qt_+\leq q<\qt_-$ && \\ \hline
(i)--(iii)& $27\leq q\leq\qmax$ & $b_2^*\leq b\leq b_1^*$ & $b_2^*<b<b_1^*$\\
(iv)--(vi) & $\qt_-\leq q\leq\qmax$ &&
\\ \hline\hline
\end{tabular}
\label{table_ER6}
\end{table}

\subsection{Class VII: $r_*\geq3$ and $\theta_2\leq\theta_*<\theta_3$}
\begin{figure}[H]
\centering
\includegraphics[width=6cm]{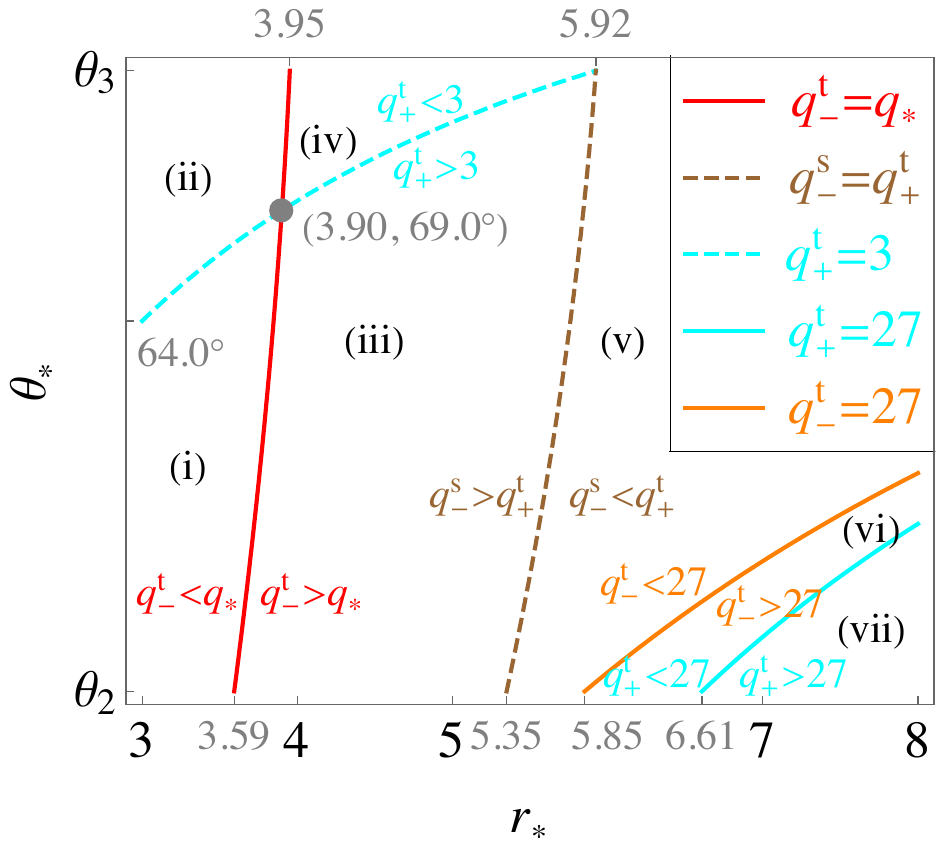}
\caption{(Class VII) Relationship between the critical $q$'s and $q_*$.}
\label{fig_q7}
\end{figure}

\begin{figure}[H]
\centering
\includegraphics[width=8cm]{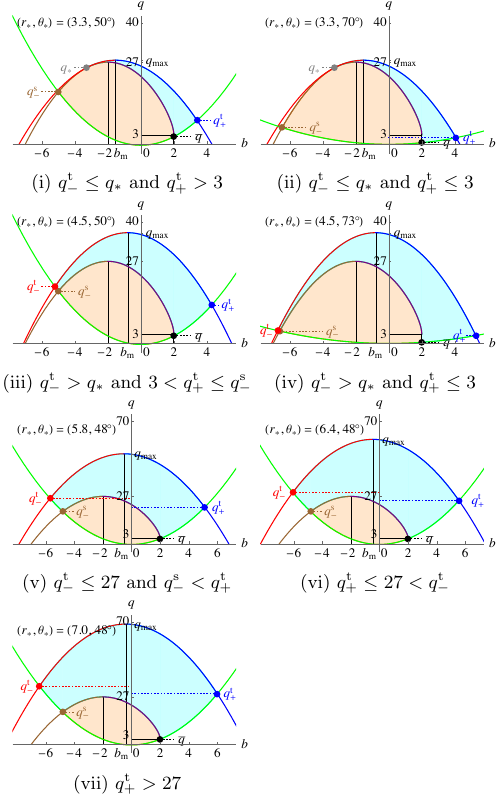}
\caption{(Class VII) Typical shape of the escapable region for $r_*\geq3$ and $\theta_2\leq\theta_*<\theta_3$. The blue, purple, brown, and green curves denote $b=b_1^*(r_*;q)$, $\bs_1(q)$, $\bs_2(q)$, and $\pm B(\theta_*;q)$, respectively. The gray, black, blue, red, and brown dashed lines denote $q=q_*(r_*)$, $\qb(\theta_*)$, $\qt_+(r_*,\theta_*)$, $\qt_-(r_*,\theta_*)$, and $\qs_-(\theta_*)$, respectively.}
\label{fig_ER7}
\end{figure}

There are seven cases according to the relative values of $q_*$, $\qt_\pm$, $\qs_-$, $3$, and $27$ (see Fig.~\ref{fig_q7}):
\begin{align}
\mathrm{(i)}&~\qt_-\leq q_* ~\mathrm{and}~ \qt_+>3,\\
\mathrm{(ii)}&~\qt_-\leq q_* ~\mathrm{and}~ \qt_+\leq3,\\
\mathrm{(iii)}&~\qt_->q_* ~\mathrm{and}~ 3<\qt_+\leq\qs_-,\\
\mathrm{(iv)}&~\qt_->q_* ~\mathrm{and}~ \qt_+\leq3,\\
\mathrm{(v)}&~\qt_-\leq27 ~\mathrm{and}~ \qs_-<\qt_+,\\
\mathrm{(vi)}&~\qt_+\leq27<\qt_-,\\
\mathrm{(vii)}&~\qt_+>27,
\end{align}
where $q_*$ contributes to specifying the escapable region for cases (i) and (ii), and $\qt_-$ contributes to that for cases (ii)--(vii). The escapable regions in the above cases are summarized in Table~\ref{table_ER7} and Fig.~\ref{fig_ER7}.

Note that solving $\qt_-(r_*,\theta_3)=27$ and $\qt_+(r_*,\theta_3)=27$ for $r_*$, we obtain $r_*\simeq19.39$ and $r_*\simeq19.61$, respectively.

\begin{table}[b]
\centering
\caption{(Class VII) Escapable region ($b,q$) for $r_*\geq3$ and $\theta_2\leq\theta_*<\theta_3$.}
\begin{tabular}{llll}
\hline\hline
Case~~ & $q$ & $b$ ($\s_r=+$) & $b$ ($\s_r=-$) 
\\ \hline
(i)--(vii)& $\qmin\leq q<\qb$ & $-B\leq b\leq B$ & n/a\\ \hline
(i), (iii), (v)--(vii) & $\qb\leq q<3$ & $-B\leq b\leq B$ & $2<b\leq B$\\
(ii) and (iv)& $\qb\leq q<\qt_+$ &&\\ \hline
(i) and (iii) & $3\leq q<\qt_+$ & $-B\leq b\leq B$ & $\bs_1<b\leq B$\\
(v)--(vii)& $3\leq q<\qs_-$ &&\\ \hline
(ii) and (iv)& $\qt_+\leq q<3$ & $-B\leq b\leq b_1^*$ & $2<b<b_1^*$\\ \hline
(i) and (iii) & $\qt_+\leq q<\qs_-$ & $-B\leq b\leq b_1^*$ & $\bs_1<b<b_1^*$\\
(ii) and (iv)& $3\leq q<\qs_-$ && \\ \hline
(i) and (ii)& $\qs_-\leq q<q_*$ & $\bs_2<b\leq b_1^*$ & $\bs_1<b<b_1^*$\\ \hline
(v) and (vi)& $\qs_-\leq q<\qt_+$ & $-B\leq b\leq B$ & $-B\leq b<\bs_2$\\
(vii)& $\qs_-\leq q<27$ && ~and $\bs_1<b\leq B$\\ \hline
(iii) and (iv)& $\qs_-\leq q<\qt_-$ & $-B\leq b\leq b_1^*$ & $-B\leq b<\bs_2$\\
(v)& $\qt_+\leq q<\qt_-$ && ~and $\bs_1<b<b_1^*$\\
(vi)& $\qt_+\leq q<27$ && \\ \hline
(i) and (ii)& $q_*\leq q<27$ & $b_2^*\leq b\leq b_1^*$ & $b_2^*<b<\bs_2$\\
(iii)--(v)& $\qt_-\leq q<27$ && ~and $\bs_1<b<b_1^*$\\ \hline
(vii)& $27\leq q<\qt_+$ & $-B\leq b\leq B$ & $-B\leq b\leq B$\\ \hline
(vi)& $27\leq q<\qt_-$ & $-B\leq b\leq b_1^*$ & $-B\leq b<b_1^*$\\
(vii)& $\qt_+\leq q<\qt_-$ && \\ \hline
(i)--(v) & $27\leq q\leq\qmax$ & $b_2^*\leq b\leq b_1^*$ & $b_2^*<b<b_1^*$\\ 
(vi) and (vii) & $\qt_-\leq q\leq\qmax$ &&
\\ \hline\hline
\end{tabular}
\label{table_ER7}
\end{table}


\newpage

\subsection{Class VIII: $r_*\geq3$ and $\theta_3\leq\theta_*<\pi/2$}
\begin{figure}[t]
\centering
\includegraphics[width=8cm]{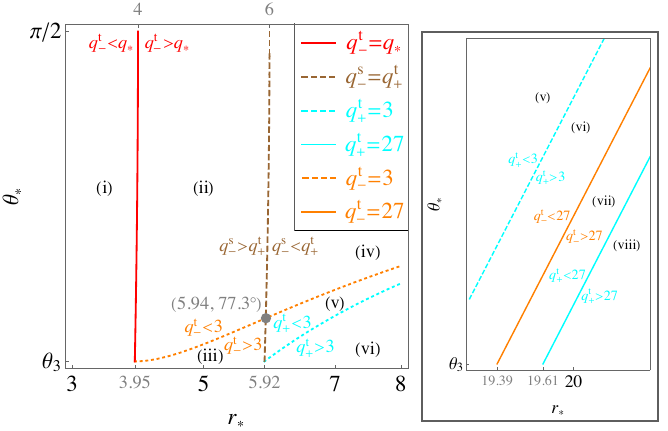}
\caption{(Class VIII) Relationship between the critical $q$'s and $q_*$.}
\label{fig_q8}
\end{figure}

There are eight cases according to the relative values of $q_*$, $\qt_\pm$, $\qs_-$, $3$, and $27$ (see Fig.~\ref{fig_q8}):
\begin{align}
\mathrm{(i)}&~\qt_-\leq q_*,\\
\mathrm{(ii)}&~q_*<\qt_-\leq3 ~\mathrm{and}~ \qs_-\geq\qt_+,\\
\mathrm{(iii)}&~\qt_->3 ~\mathrm{and}~ \qs_-\geq\qt_+,\\
\mathrm{(iv)}&~\qt_-\leq3 ~\mathrm{and}~ \qs_-<\qt_+,\\
\mathrm{(v)}&~\qs_-<\qt_+\leq3<\qt_-,\\
\mathrm{(vi)}&~\qt_+>3 ~\mathrm{and}~ \qt_-\leq27,\\
\mathrm{(vii)}&~\qt_+\leq27<\qt_-,\\
\mathrm{(viii)}&~\qt_+>27,
\end{align}
where $q_*$ contributes to specifying the escapable region only for case (i), and $\qt_-$ contributes to that for cases (i)--(viii).
The escapable regions in the above cases are summarized in Table~\ref{table_ER8} and Fig.~\ref{fig_ER8}.

\begin{table}[t]
\centering
\caption{(Class VIII) Escapable region ($b,q$) for $r_*\geq3$ and $\theta_3\leq\theta_*<\pi/2$.}
\begin{tabular}{llll}
\hline\hline
Case~~ & $q$ & $b$ ($\s_r=+$) & $b$ ($\s_r=-$) 
\\ \hline
(i)--(viii)& $\qmin\leq q<\qb$ & $-B\leq b\leq B$ & n/a\\ \hline
(i)--(iii)& $\qb\leq q<\qt_+$ & $-B\leq b\leq B$ & $2<b\leq B$\\
(iv) and (v) & $\qb\leq q<\qs_-$ && \\
(vi)--(viii) & $\qb\leq q<\qs_-$ && \\ \hline
(i)--(iii)& $\qt_+\leq q<\qs_-$ & $-B\leq b\leq b_1^*$ & $2<b<b_1^*$\\ \hline
(iv) and (v)&$\qs_-\leq q<\qt_+$ & $-B\leq b\leq B$ & $-B\leq b<\bs_2$\\
(vi)--(viii)&$\qs_-\leq q<3$ && ~and $2<b\leq B$\\ \hline
(i) & $\qs_-\leq q<3$ & $\bs_2<b\leq b_1^*$ & $2<b<b_1^*$\\
&$3\leq q<q_*$ & $\bs_2<b\leq b_1^*$ & $\bs_1<b<b_1^*$\\ \hline
(ii) & $\qs_-\leq q<\qt_-$ & $-B\leq b\leq b_1^*$ & $-B\leq b<\bs_2$\\
(iii) & $\qs_-\leq q<3$ && ~and $2<b<b_1^*$\\
(iv) & $\qt_+\leq q<\qt_-$ &&\\
(v) & $\qt_+\leq q<3$ &&\\ \hline
(ii) and (iv)&$\qt_-\leq q<3$ & $b_2^*\leq b\leq b_1^*$ & $b_2^*<b<\bs_2$\\&&& ~and $2<b<b_1^*$\\ \hline
(vi) and (vii) & $3\leq q<\qt_+$ & $-B\leq b\leq B$ & $-B\leq b<\bs_2$\\
(viii) & $3\leq q<27$&& ~and $\bs_1<b\leq B$\\ \hline
(iii) and (v)&$3\leq q<\qt_-$ & $-B\leq b\leq b_1^*$ & $-B\leq b<\bs_2$\\
(vi)&$\qt_+\leq q<\qt_-$ && ~and $\bs_1<b<b_1^*$\\
(vii)&$\qt_+\leq q<27$ && \\ \hline
(i)&$q_*\leq q<27$ & $b_2^*\leq b\leq b_1^*$ & $b_2^*<b<\bs_2$\\
(ii) and (iv)& $3\leq q<27$ && ~and $\bs_1<b<b_1^*$\\
(iii), (v), (vi)& $\qt_-\leq q<27$ && \\ \hline
(viii) &$27\leq q<\qt_+$ & $-B\leq b\leq B$ & $-B\leq b\leq B$\\ \hline
(vii)&$27\leq q<\qt_-$ & $-B\leq b\leq b_1^*$ & $-B<b<b_1^*$\\
(viii)&$\qt_+\leq q<\qt_-$ && \\ \hline
(i)--(vi) &$27\leq q\leq\qmax$ & $b_2^*\leq b\leq b_1^*$ & $b_2^*<b<b_1^*$\\ 
(vii) and (viii) &$\qt_-\leq q\leq\qmax$ &&
\\ \hline\hline
\end{tabular}
\label{table_ER8}
\end{table}

\section{DISCUSSION}
\label{sec:6}
We have completely classified the necessary and sufficient range of the impact parameters $(b,q)$ for photons emitted from an arbitrary spacetime position of the extremal Kerr black hole to escape to infinity, i.e., the escapable regions.
The radial equation of motion determines the necessary conditions for photons emitted from $r=r_*$ to escape to infinity, and the polar angle equation of motion further restricts the allowed region of photon motion.
In the process of classifying photon escape, we have defined four critical angles at which the classification of the escapable region varies qualitatively and five critical values of $q$ at which the classification of the impact parameter range varies qualitatively.
We have divided the entire spacetime into eight regions by three critical angles and $r=3$, class I--VIII.
Furthermore, we have appropriately selected the critical values of $q$ contributing to specifying the escapable region and have completely classified the difference in the shape of the escapable region, that is, the difference in the escapable parameter range, according to the relative values of critical $q$'s.
Our main results are summarized in the tables of Sec.~\ref{sec:5}.

This study has generalized our previous result~\cite{Ogasawara:2020frt}, which focused only on light sources near the horizon, to the classification that covers light sources in the entire region.
We have considered the extremal Kerr black hole here, but our classification method can be directly applied to nonextremal Kerr black holes.
Furthermore, since this method also can be applied to timelike particles, it will be possible to discuss the neutrino radiation~\cite{Wang:2021elf}, the escape of high-energy particles in high-energy astrophysics, e.g., the collisional Penrose process~\cite{Piran:1975apj,Schnittman:2018ccg}, and high-energy particle collision~\cite{Banados:2009pr,Harada:2014vka}.

As we have mentioned in the Introduction, evaluating a photon escape probability is essential to reveal the observability of phenomena around a black hole.
In the calculation of the escape probability, it is necessary to specify not only an emitter's position but also its proper motion.
However, since our complete set of the escapable regions is independent of the proper motion, the set provides a basis for evaluating the escape probability.
Based on the classification in the present paper, we will report the escape cone and probability for various states of an emitter in a forthcoming paper~\cite{Ogasawara:tbp}.

\begin{acknowledgments}
The authors are grateful to Takahiro Tanaka, Kouji Nakamura, Kazunori Kohri, and Takahiro Matsubara for useful comments.
This work was supported by
JSPS KAKENHI Grants No.~JP20J00416 and No.~JP20K14467 (K.O.) and Grant No.~JP19K14715 (T.I.).
\end{acknowledgments}

\appendix

\section{Photon escape for $\theta_*=0$ and $\theta_*=\pi/2$}
\label{App:theta-0-pi}
\subsection{$\theta_*=0$}
We consider photon escape in the case $\theta_*=0$.
When $\theta_*=0$, the regularity of the function $\Theta$ [Eq.~\eqref{def:Theta}] requires $b=0$. Substituting it into $\Theta\geq0$, we have $q\geq-1$.

We focus on the negative range $-1\leq q<0$. As shown in Sec.~\ref{subsec:3-3}, the non-negativity of the function $R$ gives the allowed parameter range of $q$.
Combining the inequality \eqref{allowed_negativeq} with $b=0$ and $-1\leq q<0$, we have
\begin{align}
-1\leq q <0<\frac{r(r^3+r+2)}{(r-1)^2}.
\end{align}
Since this inequality always holds outside the horizon, all of the photons emitted outwardly with $-1\leq q<0$ can escape to infinity.

Next, we focus on the non-negative range of $q$. In this case, $\qt_\pm$ coincide with each other and also $\qs_\pm$ coincide with each other, and their values are given by
\begin{align}
\qt_\pm(r_*,0)&=\qt_0\=\frac{r_*(r_*^3+r_*+2)}{(r_*-1)^2},\\
\qs_\pm(0)&=\qs_0\=11+8\sqrt{2}.
\end{align}

Note that $\qt_0 \geq \qs_0$ holds outside the horizon and the equal sign holds only when $r_*=1+\sqrt{2}$.
There are two cases depending on the radial position of the emitter: 
\begin{align}
\mathrm{(i)}&~ r_*\leq 1+\sqrt{2},\\
\mathrm{(ii)}&~ r_*>1+\sqrt{2}.
\end{align}
The escapable regions in the above cases are summarized in Table \ref{table_ERtheta0}.

\newpage

\begin{figure}[H]
\centering
\includegraphics[width=8cm]{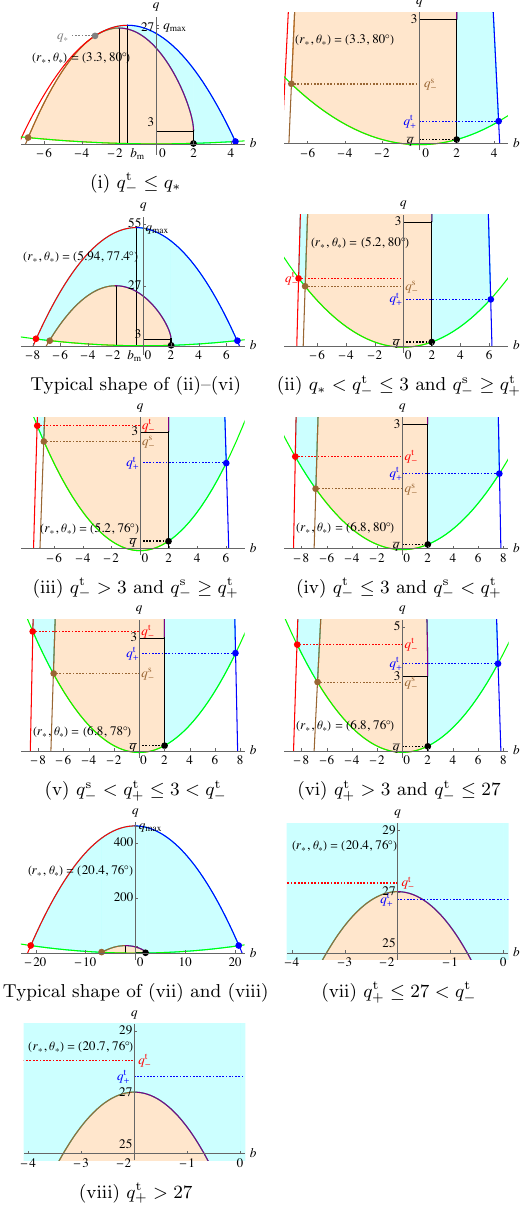}
\caption{(Class VIII) Typical shape of the escapable region for $r_*\geq3$ and $\theta_3\leq\theta_*<\pi/2$. The blue, purple, brown, and green curves denote $b=b_1^*(r_*;q)$, $\bs_1(q)$, $\bs_2(q)$, and $\pm B(\theta_*;q)$, respectively. The gray, black, blue, red, and brown dashed lines denote $q=q_*(r_*)$, $\qb(\theta_*)$, $\qt_+(r_*,\theta_*)$, $\qt_-(r_*,\theta_*)$, and $\qs_-(\theta_*)$, respectively.}
\label{fig_ER8}
\end{figure}

\begin{table}[t]
\centering
\caption{Escapable region ($b,q$) for $\theta_*=0$.}
\begin{tabular}{llll}
\hline\hline
Case~~ & $q$ & $b$ ($\s_r=+$) & $b$ ($\s_r=-$) 
\\ \hline
(i) & $-1\leq q<\qs_0$ & $b=0$ & n/a\\ 
& $q\geq \qs_0$ & n/a & n/a \\ \hline
(ii) & $-1\leq q\leq\qs_0$ & $b=0$ & n/a\\
& $\qs_0< q<\qt_0$ & $b=0$ & $b=0$\\
& $q=\qt_0$ & $b=0$ & n/a\\
& $q>\qt_0$ & n/a & n/a
\\ \hline\hline
\end{tabular}
\label{table_ERtheta0}
\end{table}

\subsection{$\theta_*=\pi/2$}
In the case of $\theta_*=\pi/2$, the non-negativity of $\Theta$ reads $q\geq0$. Therefore, the necessary parameter regions for photon escape in Table \ref{table_necessary} are identified with the escapable region. The corresponding figures, i.e., the escapable region for $\theta_*=\pi/2$, are found in Fig.~\ref{fig_necessary}.

\section{Equation for the polar angle of Kerr geodesics}
\label{App:ThetaEq}

We focus on the function $\Theta$, which appears in the geodesic equation for the polar angle direction of the Kerr spacetime. 
We consider the following equation: 
\begin{align}
\label{eq:THETA1}
\Theta=q-b^2\cot^2\theta+a^2\cos^2\theta=0,
\end{align}
where $a (>0)$, $b$, $q$ are constants, and $0\le \theta \le \pi$. 
For $\theta=0, \pi$, the constant $b$ must vanish, and $q=-a^2$ must hold. 
We assume $0< \theta < \pi$ in what follows. 
Equation~\eqref{eq:THETA1} is rewritten as an equivalent equation in terms of $\sin \theta$,
\begin{align}
\label{eq:THETA2}
a^2 \sin^4\theta -(a^2+b^2+q) \sin^2\theta+b^2=0.
\end{align}
Solving Eq.~\eqref{eq:THETA2} for $\sin^2\theta$, we obtain 
\begin{align}
\label{eq:THETAsin2}
\sin^2\theta
&=\frac{1}{2a^2}\left[\:\!
a^2+b^2+q\pm
\sqrt{(a^2+b^2+q)^2-4a^2b^2}
\:\!\right].
\end{align}
In order for $\sin^2\theta$ to be real,  the parameters must satisfy the inequality
\begin{align}
\label{eq:THETAsin0}
q+\left(
|\:\!b\:\!|-a
\right)^2\ge 0,
\end{align}
which also guarantees $\sin^2\theta$ positive.
On the other hand, the condition $\sin^2\theta\le 1$ is written as
\begin{align}
\label{eq:THETAsin2sm1}
a^2-b^2-q\mp \sqrt{(a^2-b^2-q)^2+4a^2 q} \ge 0,
\end{align}
where the double sign corresponds to that in Eq.~\eqref{eq:THETAsin2}.
For the upper case, the parameters must satisfy $q\leq 0$ and $|b|\leq a$.
For the lower case, if $a^2-b^2-q\ge 0$ together with Eq.~\eqref{eq:THETAsin0}, then the inequality~\eqref{eq:THETAsin2sm1} holds; if $a^2-b^2-q< 0$, 
 then $q\geq 0$ must hold. 

Now we introduce new combinations of the parameters
\begin{align}
\zeta_\pm=\sqrt{(b\pm a)^2+q},
\end{align}
which satisfy the following relations: 
\begin{align}
\zeta_+\zeta_-&=\sqrt{(a^2+b^2+q)^2-4a^2b^2},
\\
\zeta_+^2+\zeta_-^2&=2(a^2+b^2+q),
\\
\zeta_+^2-\zeta_-^2&=4ab.
\end{align}
Using these, we can rewrite Eq.~\eqref{eq:THETAsin2} in terms of $\zeta_\pm$ as
\begin{align}
\sin^2\theta=\frac{(\zeta_+\pm \zeta_-)^2}{4a^2}.
\end{align}
Because of the range of $\theta$, we can take the positive branch
\begin{align}
\sin \theta=\frac{|\:\!\zeta_+\pm \zeta_-\:\!|}{2a}.
\end{align}
Finally we obtain 
\begin{widetext}
\begin{align}
\label{eq:THETAsinsol}
\sin \theta=
\left\{
\begin{array}{lll}
\dfrac{\zeta_++\zeta_-}{2a} & \mathrm{for} & |b|\le a, \ 
-(|\:\!b\:\!|-a)^2\le q\leq 0,
\\[3mm]
\dfrac{\zeta_+-\zeta_-}{2a} & \mathrm{for} & 
\left[\:\!
0\le b<a, \ q\ge -(b-a)^2\:\!\right] \ \mathrm{or} \ 
\left[\:\!
b\ge a, \ q\ge 0
\:\!\right],
\\[3mm]
\dfrac{\zeta_--\zeta_+}{2a} & \mathrm{for} & 
\left[\:\!
-a<b<0, \ q\ge -(b+a)^2
\:\!\right]\ \mathrm{or}
\ 
\left[\:\!
b\leq -a, \ q>0
\:\!\right].
\end{array}
\right.
\end{align}
\end{widetext}


\end{document}